%% file: main.tex
\documentclass[aps,prx,reprint,superscriptaddress,floatfix]{revtex4-2}

\usepackage{amsmath}
\usepackage{amssymb}
\usepackage[linesnumbered,ruled,vlined]{algorithm2e}
\usepackage{graphicx}
\usepackage{overpic}
\usepackage[caption=false,position=top,justification=raggedright]{subfig}

\usepackage{dcolumn}
\usepackage{bm}
\usepackage{hyperref}
\usepackage{xcolor}
\usepackage{soul}
\usepackage{threeparttable,tablefootnote,booktabs}
\usepackage{tikz}
\usepackage{placeins}

\newcommand{\Hamil}{\hat{H}}
\newcommand{\bra}[1]{\langle #1 |}
\newcommand{\ket}[1]{| #1 \rangle}
\newcommand{\braket}[2]{\langle #1 | #2 \rangle}

\begin{document}


\title{Large-scale stochastic propagation method beyond the sequential approach}

\author{Zhichang Fu}
\thanks{These authors contributed equally to this work.} 
\affiliation{Key Laboratory of Artificial Micro- and Nano-structures of Ministry of Education and School of Physics and Technology, Wuhan University, Wuhan 430072, China}

\author{Yunhai Li}
\thanks{These authors contributed equally to this work.} 
\affiliation{Wuhan Institute of Quantum Technology, Wuhan 430206, China}%

\author{Weiqing Zhou}
\email{weiqingzhou@whu.edu.cn} 
\affiliation{Key Laboratory of Artificial Micro- and Nano-structures of Ministry of Education and School of Physics and Technology, Wuhan University, Wuhan 430072, China}%
\affiliation{Wuhan Institute of Quantum Technology, Wuhan 430206, China}%

\author{Shengjun Yuan}
\email{s.yuan@whu.edu.cn} 
\affiliation{Key Laboratory of Artificial Micro- and Nano-structures of Ministry of Education and School of Physics and Technology, Wuhan University, Wuhan 430072, China}%
\affiliation{Wuhan Institute of Quantum Technology, Wuhan 430206, China}%
\affiliation{School of Artificial Intelligence, Wuhan University, 430072, Wuhan, China}%

\date{\today}

\begin{abstract}
The $O(N)$ stochastic propagation method, which relies on the numerical solution of the time-dependent Schr\"odinger equation using random initial states, is widely used in large-scale first-principles calculations. In this work, we eliminate the conventional sequential computation of intermediate states by introducing a concurrent strategy that minimizes information redundancy. The new method, in its state-, moment-, and energy-based implementations, not only surpasses the time step constraint of sequential propagation but also maintains precision within the framework of the Nyquist-Shannon sampling theorem. Systematic benchmarking on one billion atoms within the tight-binding model demonstrates that our new concurrent method achieves up to an order-of-magnitude speedup, enabling the rapid computation of a wide range of electronic, optical, and transport properties. This performance breakthrough offers valuable insights for enhancing other time-propagation algorithms, including those employed in large-scale stochastic density functional theory.
\end{abstract}

\maketitle

\section{Introduction}
\input{context/introduction}

\section{Methods}
\label{sec:methods}

\subsection{The Concurrent Stochastic Propagation Method}
\label{sec:sPM}
\input{context/sPM}

\subsection{Optimal Block Size Selection}
\label{subsec:block_selection}
\input{context/block_selection}

\section{Density of States and Local Density of States}
\label{sec:DOS}
\input{context/DOS}

\section{Quasi-eigenstates}
\label{sec:QE}
\input{context/QE}

\section{electronic conductivity}
\label{sec:DC}
\input{context/DC}

\section{Optical Conductivity}
\label{sec:OC}
\input{context/OC}

\section{Dynamic Polarization}
\label{sec:DP}
\input{context/DP}

\section{Charge Density}
\label{sec:CD}
\input{context/CD}







\section{Conclusion}
\label{sec:conclusion}
\input{context/conclusion}

\begin{acknowledgments}

This work was supported by the National Natural Science Foundation of China (Grants No. 12425407 and 12174291) and the Major Program (J.D.) of Hubei Province (Grant No. 2023BAA020). W.Z. gratefully acknowledges support from the Hongyi postdoctoral fellowship of Wuhan University. We thank the Core Facility of Wuhan University for providing the computational resources.
\end{acknowledgments}

\appendix

\section{The analysis of cost and memory}
\label{sec:mem}
\input{context/cost_mem_table}

\section{Details of polynomial expansion methods}
\label{sec:polys}
\input{context/poly_expansions}





\FloatBarrier
\bibliography{refs.bib}

\end{document}

%% file: context/introduction.tex
Numerical simulations based on quantum mechanics often boil down to solving the Schrödinger equation in various forms through numerical methods. Matrix diagonalization represents one of the most straightforward algorithms for addressing this type of generalized eigenvalue problem. For instance, in first-principles calculations based on Kohn–Sham density functional theory~\cite{1964-hk,1965-ks,1999-Goedecker}, eigenfunctions and their corresponding eigenvalues are obtained directly via diagonalization. These eigenstates are then used to reconstruct the electron density during the self-consistent field iterations~\cite{2014-ELPA}. However, the computational cost scales cubically with system size, making it one of the major bottlenecks in large-scale simulations including defects~\cite{2014-defects}, alloys~\cite{1990-alloys}, fractals~\cite{2013-fractal}, quasicrystals~\cite{1984-quasicrystal}, clusters~\cite{1993-cluster}, interface~\cite{1958-cahn}, heterostructures~\cite{2013-geim}, amorphous structures~\cite{2011-Berthier} and large molecules such as DNA~\cite{2000-DNA}. Traditional methods struggle to handle these complex quantum systems effectively.

Developing linear-scaling quantum simulation algorithms has become crucial to addressing this challenge. Strategies such as fragment-based methods~\cite{2012-Gordon-Fragmentation-review}, polynomial expansion~\cite{1999-Goedecker}, direct optimization~\cite{2012-bowler-review}, matrix purification~\cite{1960-McWeeny-RMP}, and stochastic techniques~\cite{1989-hutchinson-stochastic,2000-hams-eigenvalues} provide robust mathematical tools for achieving this goal. Since the 1990s, these applied mathematical approaches have been progressively integrated into quantum simulations. Leveraging the locality principle in quantum mechanics~\cite{1996-kohn}, a number of linear-scaling methods—also referred to as $O(N)$ methods—have been developed~\cite{1991-yang-dc,1995-yang-dm-dc,2013-sdft,1993-LNV,1993-Mauri-prb,1994-Mauri-electronic,1994-Goedecker-FOE,2002-Niklasson-expanson,2006-weisse-kpm,2006-ozaki-DC-Krylov}. Their computational load increases only linearly with system size, thereby extending the reach of quantum mechanical accuracy to larger scales~\cite{1999-Goedecker,2012-bowler-review,2015-akimov-large,2019-lin-numerical}.

Stochastic propagation methods (sPM) is a class of linear-scaling approaches that combine Fourier spectral analysis with stochastic techniques~\cite{2000-hams-eigenvalues,2010-yuan-graphene,2012-time-domain-MP2,2013-sGW,2015-BSE,2023-dfpm}. It transforms the solution of the stationary Schr\"odinger equation into the evolution of stochastic states governed by the time-dependent Schr\"odinger equation (TDSE). Physical quantities can be accurately determined by analyzing the temporal correlations of these wavefunctions.
Crucially, due to the statistical properties of random states, the error in global observables scales as $O(1/\sqrt{N})$ where $N$ denotes the number of atoms in the system. This scaling behavior demonstrates a sublinear computational cost with respect to system size. For ultra-large systems approaching the thermodynamic limit ($N\to\infty$), even a single random state can provide results with the desired accuracy. This framework enables the calculation of a wide range of physical quantities of systems with billions of atoms, including the density of states (DOS)~\cite{2000-hams-eigenvalues,2010-yuan-graphene}, local density of states (LDOS)~\cite{2020-shi-nc}, quasi-eigenstates (QE)~\cite{2010-yuan-graphene,2020-shi-nc}, optical conductivity (OC)~\cite{2010-yuan-graphene}, electronic conductivity (EC)~\cite{2010-yuan-graphene}, and dynamic polarization (DP)~\cite{2012-yuan-screening,2011-yuan-plasmons}. Based on these six fundamental quantities, numerous properties can be derived, such as carrier velocity~\cite{2010-yuan-graphene}, mobility~\cite{2010-yuan-graphene}, mean free path~\cite{2010-yuan-graphene}, localization length~\cite{2012-yuan-screening}, diffusion coefficient~\cite{2010-yuan-graphene}, response functions~\cite{2011-yuan-plasmons}, dielectric constant~\cite{2011-yuan-plasmons}, transmission coefficient~\cite{2015-logemann-klein}, energy loss spectrum~\cite{2012-yuan-screening}, plasmon spectrum~\cite{2011-yuan-plasmons}, and plasmon lifetime~\cite{2011-yuan-plasmons}.
These Tight-Binding Propagation Methods (TBPM) have been integrated into an open-source large-scale tight-binding calculation software package TBPLaS~\cite{2023-tbplas}. 

Beyond empirical models, the sPM strategy has also been extended to first-principles methods such as the second order Møller-Plesset (MP2)~\cite{2012-time-domain-MP2}, GW~\cite{2013-sGW}, and the Bethe-Salpeter equation (BSE)~\cite{2015-BSE}. Moreover, time propagation has been found to effectively suppress stochastic errors arising from non-degenerate state coupling during charge density calculations~\cite{2023-dfpm}, which explains the observation that the number of stochastic states needed in time-dependent stochastic density functional theory (td-sDFT)~\cite{2015-td-sdft} is far smaller than that in static sDFT~\cite{2013-sdft}. 
By further incorporating this approach into time-independent density functional theory, time propagation has emerged as a general strategy for reducing errors in charge density computations~\cite{2023-dfpm}.
Currently, sPM have been widely applied to simulations of various complex quantum systems~\cite{2023-hu-science,2022-long,2021-wu,2019-yu-quasi,2023-yao-fractal,2022-wang-prx,2022-liu-1D}.

In sPM, achieving a desired energy resolution through Fourier transformation from the time domain requires a sequence of step-by-step propagations. The maximum allowable time step in this process is constrained by the bandwidth of the energy spectrum. Typically, the time evolution of a quantum state is implemented numerically by expanding the time-evolution operator in a basis of orthogonal polynomials—an approach valued for its rapid convergence and unconditional stability~\cite{1939-orthogonal}.
A notable feature of such polynomial expansions is their characteristic scaling behavior: the required truncation order per unit time decreases significantly as the time step increases. This suggests that numerical efficiency would benefit from using larger time steps. However, this conflicts with the Nyquist-Shannon sampling theorem~\cite{1949-shannon-sampling}, which require wavefunction information at numerous finely spaced intermediate time points.
To resolve this fundamental limitation, we introduce an efficient strategy by thoroughly compressing the computational redundancy in the conventional sequential propagation. Our method computes directly the final state via a single, long-time propagation step and concurrently reconstructs the contributions from all necessary intermediate states during the propagation. Furthermore, an adaptive time-blocking scheme can be incorporated to optimally balance computational cost and memory usage. The proposed method appears to break the step size constraint of the Nyquist-Shannon sampling theorem~\cite{1949-shannon-sampling}, yet it ingeniously maintains precision consistency through final-state reconstruction, offering a general and efficient framework for accelerating sPM calculations in quantum systems.

\begin{figure*}[htbp]
    \centering
    \begin{overpic}[height=0.185\textheight]{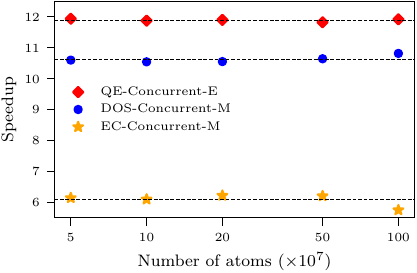}
        \put(-2,61){\bfseries (a)}
    \end{overpic}
    \quad
    \begin{overpic}[height=0.188\textheight]{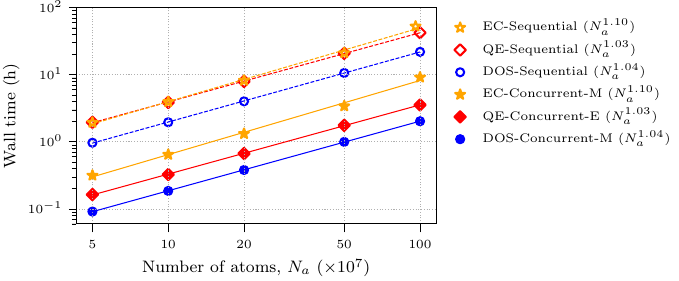}
        \put(-1,38){\bfseries (b)}
    \end{overpic}
    
    \caption{%
    \textbf{Comparison of scalability and computational performance for the new concurrent and the conventional sequential sPM in large-scale systems.}
    (a) Speedup as a function of the number of atoms ($N_a$) for density of states (DOS), electronic conductivity (EC) and quasi-eigenstates (QE). The horizontal dashed lines indicate the average speedup for each corresponding data series. Note that for the electronic conductivity, the data point for the largest system is rescaled from a calculation on $9.6 \times 10^8$ atoms, which was the maximum size feasible due to memory limitations.
    (b) Wall time as a function of the number of atoms, $N_a$. The plotted lines represent the power-law fits to the corresponding data, with the scaling exponent $c$ provided in the legend in the form of $N_a^c$.
    These tests utilized nearest-neighbor, single-layer graphene models~\cite{1947-Wallace} with system sizes spanning from 50 million to 1 billion atoms. The calculations were executed on a  computational node of 512 GB of RAM, equipped with two Intel\textregistered{} Xeon\textregistered{} Gold 6326 processors (32 cores total).
    }
    \label{fig:scalability}
\end{figure*}

As illustrated schematically in Fig.~\ref{fig:scalability}, the new concurrent stochastic propagation method delivers substantial performance improvements across a range of property calculations, with speedups exceeding an order of magnitude for DOS and quasi-eigenstate calculations, 5-6$\times$ for electronic conductivity—all without loss of accuracy compared to conventional sequential propagation. It achieves consistent acceleration across system sizes while maintaining linear scaling, reducing simulation times for billion-atom tight-binding systems on a single cluster node from days to hours.

This rest of the paper is organized as follows. Sec.~\ref{sec:sPM} revisits the sPM framework and subsequently discusses the new concurrent stochastic propagation method proposed in this work. Sec.~\ref{subsec:block_selection} gives a theoretical analysis for optimal block size selection. 
The practical application and performance of our new concurrent method are then detailed in Secs.~\ref{sec:DOS}--\ref{sec:CD}, where we evaluate its effectiveness for several key physical quantities: (local) density of states, quasi-eigenstates, electronic conductivity, optical conductivity, dynamical polarization, and charge density. For each of these quantities, we first introduce its theoretical foundation and implementation details, then validate its numerical accuracy against the sequential sPM, and finally conduct a comprehensive performance evaluation. This evaluation quantitatively benchmarks computational efficiency and memory requirements and provides guidelines for optimal parameter selection.
Finally, Sec.~\ref{sec:conclusion} summarizes the key findings and conclusions of this work.

%% file: context/sPM.tex
Within the stochastic propagation method, physical quantities are derived from the time evolution of random states without the need for matrix diagonalization. The requisite mathematical expressions can be classified into three types. The first type concerns the computation of a wave function at a given energy, formulated as:
\begin{equation}
\ket{\Psi(E)} =  \int_{-\infty}^{\infty}  a(E, t)\mathrm{e}^{-\mathrm{i}\hat H t}\ket{\psi_1} \, \mathrm{d}t,
\label{eq:form1}
\end{equation}
and other two types are associated with correlation functions incorporating either one or two time-evolution operators:

\begin{equation}
A_1(E) = \int_{-\infty}^{\infty} a(E, t) \bra{\psi_1}\mathrm{e}^{-\mathrm{i}\hat H t}\ket{\psi_2} \, \mathrm{d}t,
\label{eq:form2}
\end{equation}
\begin{equation}
A_2(E) = \int_{-\infty}^{\infty} a(E, t) \bra{\psi_1} \mathrm{e}^{\mathrm{i}\hat H t}\hat O \mathrm{e}^{-\mathrm{i}\hat H t}\ket{\psi_2} \, \mathrm{d}t.
\label{eq:form3}
\end{equation}
Here $a(E, t)$ denotes an energy- and time-dependent coefficient and $\mathrm{e}^{-\mathrm{i}\hat H t}$ is the time-evolution operator. The wave functions $\ket{\psi_1}$ and $\ket{\psi_2}$ are introduced using a single initial random state $\ket{\psi}$ as $\ket{\psi_1}=\hat{O_1}\ket{\psi}$ and $\ket{\psi_2}=\hat{O_2}\ket{\psi}$. These operators, $\hat O$, $\hat O_1$ and $\hat O_2$ are time-independent operators, including but not limited to the identity operator $\hat I$, the Fermi-Dirac operator $f(\hat H)$, the current density operator $\hat J$, the density operator $\hat\rho(\mathbf{q})$ and the real-space projection operator $\ket{\mathbf{r}}\bra{\mathbf{r}}$.

In the common approach of sPM, the propagation of a state in the time domain is performed step-by-step using a fixed time step $\tau$
\begin{equation}
  \ket{\psi(t_j)} = \mathrm{e}^{-\mathrm{i} \hat{H}\tau}|\psi(t_{j-1})\rangle,
\end{equation}
where $t_j=j\cdot \tau$, to sequentially obtain a series of states with equally spaced time, $\{\ket{\psi(t_1)},\ket{\psi(t_2)},...,\ket{\psi(t_{N_t})}\}$. Here $N_t$ is the total number of time steps which determines the energy resolution of the calculated quantities in Eq.~\ref{eq:form1}-\ref{eq:form3}. 

The decomposition of the time-evolution operator $\mathrm{e}^{-\mathrm{i}\hat H t}$ is numerically stable and accurate to use an orthogonal polynomial expansion~\cite{1939-orthogonal}. Commonly used polynomial forms include Chebyshev polynomials~\cite{2021-jin-random,2006-weisse-kpm}, Legendre polynomials~\cite{2016-lin-poly,2010-qin-poly}, and the more general Jacobi polynomials~\cite{2011-qin-poly,2025-poly}. These expansion methods impose no special requirements on the form of the Hamiltonian and are applicable to arbitrarily complex structures, but they require the eigenvalues of the Hamiltonian matrix to be rescaled to lie within the interval $[-1,1]$. Generalized Laguerre polynomials~\cite{2011-spectral} and Hermite polynomials~\cite{2010-qin-poly,2011-qin-poly} are suitable for spectra with an exponential decay profile on $[0,\infty)$ and for those with a Gaussian profile over the entire real axis, such as in calculations involving quantum scattering and quantum harmonic oscillators. All of these polynomials can be generated to higher orders using three-term recurrence relations; the formal details are provided in Appendix~\ref{sec:polys}.


The Chebyshev expansion, owing to its advantageous minimax property, rapid convergence, and unconditional numerical stability, is one of the most prominent polynomial methods. Mathematically, The time-evolution operator with a time step $\tau$ can be expressed as an expansion in Chebyshev operators $T_n(H)$ ~\cite{2021-jin-random}:
\begin{equation}
  \mathrm{e}^{-\mathrm{i}\tilde{H} \tilde{\tau}} = \sum_{n=0}^{N(\tilde{\tau})} c_n(\tilde{\tau}) T_n(\tilde{H}),
\end{equation}
Here, all symbols with a tilde denote rescaled quantities to meet the requirement of confining the energy spectrum within $[-1,1]$ for Chebyshev expansion. $\tilde{H} = (\hat H-H_0)/|H|$, where $H_0$ and $|H|$ are the spectral center and half-width of the input Hamiltonian $\hat H$, respectively, which are typically obtained using the Lanczos algorithm~\cite{1950-lanczos-iterative}. $\tilde{\tau} = |H| \cdot \tau$ is the rescaled time step and $N(\tilde{\tau})$ is the corresponding polynomial truncation order, and $T_n(\tilde{H})$ is the Chebyshev polynomial of the first kind, which can be obtained recursively. In practice, the propagation is implemented by recursively generating the Chebyshev states $T_n(\tilde{H})|\psi\rangle$ via matrix-vector multiplications, rather than through direct matrix-matrix multiplication. The primary computational cost is proportional to the total number of time steps, $N_t$, and the number of expansion terms per step, $N(\tilde{\tau})$. 
%
%
%
%
According to the Nyquist-Shannon sampling theorem~\cite{1949-shannon-sampling}, the time step $\tau$ is determined by the spectral width in the energy domain, $\Delta E$, such that $\tau_{\mathrm{NS}} = 2\pi / \Delta E$, to avoid spectral aliasing. The spectral width of a physical quantity is typically equal to the spectral width of the Hamiltonian, i.e., $\Delta E = 2|H|$. Consequently, the rescaled time step in the sequential approach of sPM is fixed as $\tilde{\tau}_{\mathrm{NS}} = \pi$.

To set the stage for the new concurrent approach, we first examine the computational efficiency of varying the time step under a fixed total propagation duration. Here, we introduce the concept of expansion load $R(\tilde{\tau}) = N(\tilde{\tau})/\tilde{\tau}$, where the number of polynomial truncation order $N(\tilde{\tau})$ is associated with a given rescaled time step $\tilde{\tau}$. The expansion load $R(\tilde{\tau})$ quantifies the number of polynomials computed per unit time for a given time step. A higher value indicates lower computational efficiency, as it necessitates calculating more polynomial orders to achieve the same total propagation duration; conversely, a lower value signifies higher efficiency. In practice, the number of polynomial truncation order, $N(\tilde{\tau})$, is determined by a predefined precision threshold $\eta$, such that the expansion coefficients $c_{n}(\tilde{\tau})$ satisfy $|c_{n}(\tilde{\tau})| > \eta$ for all $n < N(\tilde{\tau})$. The threshold is typically set to $\eta=10^{-14}$ in double-precision computations, a value chosen to avoid numerical artifacts while being close to the machine epsilon.

\begin{figure*}[htbp]
        \begin{overpic}[width=0.47\textwidth]{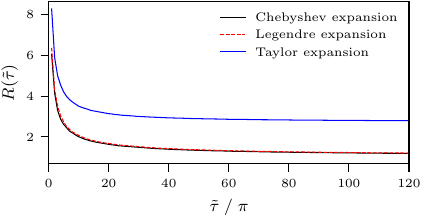}
        \put(-2,48){\bfseries (a)}
    \end{overpic}
    \quad
    \begin{overpic}[width=0.46\textwidth]{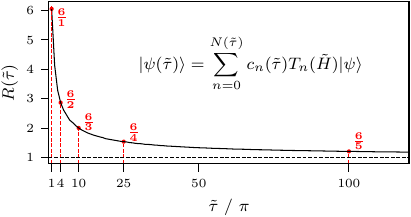}
        \put(-1,49.5){\bfseries (b)}
    \end{overpic}
    \caption{
    \textbf{The Numerical Foundations of Concurrent Stochastic Propagation Method.}
    (a) Expansion load $R(\tilde{\tau}) = N(\tilde{\tau})/\tilde{\tau}$ (the number of polynomial decomposition per unit of time) vs. the rescaled time step $\tilde{\tau}$ for Chebyshev, Legendre, and Taylor expansions of the time-evolution operator $\mathrm{e}^{-\mathrm{i}\tilde{H}\tilde{\tau}}$.
    (b) $R(\tilde \tau)$ vs. $\tilde \tau$ for the Chebyshev expansion, with $\tilde \tau = \pi$ as a reference. Points are marked where $R(\tilde \tau)$ is reduced to $\tfrac12$, $\tfrac13$, $\tfrac14$, and $\tfrac15$ of the reference value.
    }
    \label{fig:overview}
\end{figure*}

 In Fig.~\ref{fig:overview}(a), we compare the expansion load $R(\tilde{\tau})$ as a function of rescaled time step $\tilde{\tau}$ for three types of polynomials. For all polynomials considered, $R(\tilde{\tau})$ first decreases rapidly with increasing time step before gradually leveling off. In other words, for a fixed propagation duration, the total number of terms required for the polynomial expansion decreases rapidly as the time step increases. For the Taylor expansion, $R(\tilde{\tau})$ decrease to a larger  value comparing to the other two expansions. Furthermore, the corresponding expansion coefficients, $(\mathrm{i}\tilde{\tau})^n/n!$, are highly unstable due to the factorial term; for large $\tilde{\tau}$, the coefficients first grow to be very large before rapidly decreasing, making the Taylor expansion unsuitable for practical calculations. The Chebyshev and Legendre expansions, in contrast, exhibit very similar behavior.

Taking the Chebyshev expansion as an example, we quantitatively analyze the relationship between the expansion load and the time step. As shown in Fig.~\ref{fig:overview}(b), the expansion load $R(\tilde{\tau})$ is approximately 6 when the time step is $\tilde{\tau}=\pi$. As the time step increases, $R(\tilde{\tau})$ rapidly decreases and approaches 1. Specifically, when the rescaled time steps are $4\pi$, $10\pi$, $25\pi$, and $100\pi$, the load $R(\tilde{\tau})$ decreases to approximately $1/2, 1/3, 1/4,$ and $1/5$ of the value at $\tilde{\tau}=\pi$, respectively. Given a fixed total propagation duration $t_{\mathrm{tot}}$, the computational cost, dictated primarily by the total number of polynomial expansions and their associated matrix-vector multiplications, decreases with the use of a larger time step $\tilde{\tau}$. However, simply increasing the time step would lead to coarse time-domain sampling, violating the Nyquist-Shannon sampling theorem and causing spectral aliasing.

We propose an efficient, concurrent propagation strategy that not only employs large time steps beyond the Nyquist-Shannon limit but also enables the parallel reconstruction of arbitrarily refined results from a single final-state propagation.
The efficacy of this approach stems from a key insight: for a given initial state $\ket{\psi}$, the propagated states $\ket{\psi(t)}$ at different times $t$ are all linear combinations of the same underlying set of Chebyshev states, $T_n(\tilde{H})\ket{\psi}$. 
Therefore, one can first perform a single propagation over the total propagation duration, $t_{\mathrm{tot}}=N_t \cdot \tau$, to compute the final state $\ket{\psi(t_{\mathrm{tot}})}$ and its corresponding $N(\tilde{t}_{\mathrm{tot}})$ Chebyshev states. These states contain all the information necessary to reconstruct any intermediate state $\ket{\psi(t_j)}$ by simply taking linear combinations of the appropriate subset of states. This procedure replaces computationally expensive matrix-vector multiplications with lower-cost scalar multiplications and vector additions.

The most direct way to implement this
is to store all $N(\tilde{t}_{\mathrm{tot}})$ Chebyshev states of the longest propagation, and reused them for the state at any intermediate time step. While straightforward, this method incurs a significant memory overhead proportional to $O(N(\tilde{t}_{\mathrm{tot}}) \cdot N_{\mathrm{basis}})$, where $N_{\mathrm{basis}}$ is the size of the basis set. For large-scale calculations, this memory cost can become a new computational bottleneck. To overcome this limitation, several more sophisticated techniques can be employed as follows.


\paragraph*{State-based Implementation:} 
In the sequential sPM, the state $\ket{\psi(t_{j})}$ at time $t_{j}$ is propagated from  $t_{j-1}$ as 
\begin{align}
   \ket{\psi(t_{j})}
    &=\sum_{n=0}^{N(\tilde{\tau})}c_n(\tilde{\tau})T_n(\tilde{H})\ket{\psi(t_{j-1})}.
\end{align}

In the new state-based implementation, one calculates and saves the intermediate propagated states $\ket{\psi(t_j)}$ in parallel. 
It enables the concurrent calculation of $\ket{\psi(t_j)}$ by reusing the shared Chebyshev states $T_n(\tilde{H})\ket{\psi}$ from a single propagation to the final state:
\begin{equation}
    \ket{\psi(t_j)} = \sum_{n=0}^{N(j\cdot\tilde{\tau})}c_n(\tilde{t}_j)T_n(\tilde{H})\ket{\psi}.
     \label{eq:TE}
\end{equation}
Comparing to save all the Chebyshev states of the longest propagation, this new state-based approach reduces the memory overhead by a factor of $R(\tilde{\tau})\cdot\pi$ to $O(N_t \cdot N_{\mathrm{basis}})$, where $N_t$ is the number of time steps. 
At this point, the propagated states at distinct time instances are computed concurrently to be utilized in subsequent calculations of physical properties. The associated additional memory overhead can be regulated by employing an adaptive time-blocking scheme which we will discuss later. The state-based implementation can be used for all calculations expressed in Eq.~\ref{eq:form1}-\ref{eq:form3}.
\paragraph*{Moment-based Implementation:} In the sequential sPM, the calculation of $A_1(E)$  in Eq.~\ref{eq:form2} is formulated as follows:
\begin{equation}
\begin{split}
A_1(E) &=  \int_{-\infty}^{\infty} a(E, t) \bra{\psi_1}  \mathrm{e}^{-\mathrm{i}\hat H t}\ket{\psi_2} \, \mathrm{d}t\\
&\approx\tau \; a(E, 0) \braket{\psi_1}{\psi_2} \\
&+\tau \sum_{j=1}^{N_t} \sum_{n=0}^{N(\tilde{\tau})}\big(a(E, t_j)\bra{\psi_1}  c_n(\tilde{\tau}) T_n(\tilde{H}) \ket{\psi_2(t_{j-1})} \\
& + a(E, -t_{j})\bra{\psi_1}  c_n(-\tilde{\tau}) T_n(\tilde{H}) \ket{\psi_2(-t_{j-1})}\big).
\label{eq:form2_sequential} 
\end{split}
\end{equation}
In this expression, the summation over the polynomial index $n$ is nested within the loop over the time index $j$. This structure necessitates repeated, time-consuming matrix-vector multiplications involving $T_n(\tilde{H})\ket{\psi_2}$, that must be performed independently for the forward and backward time propagation.
By contrast, in the new moment-based implementation, we interchange the order of summation for the time index $j$ and the polynomial index $n$, and rearrange the formula as:
\begin{equation}
A_1(E) = \sum_{n=0}^{N(\tilde{t}_{\mathrm{tot}})} m_n C_n(E),
\end{equation}
where $m_n$ is introduced as the Chebyshev moments
\begin{equation}
    m_n=\bra{\psi_1} T_n(\tilde{H})\ket{\psi_2}
\end{equation}
and the coefficient $C_n(E)$ is
\begin{equation}
\begin{split}
    C_n(E) = \tau \sum_{j=N_0}^{N_t} & \frac{(2-\delta_{j,0})}{2} \big( a(E, t_j) c_n( \tilde{t}_j) \\
    & + a(E, -t_{j}) c_n( -\tilde{t}_{j}) \big).
\end{split}
\label{eq:coeff}
\end{equation}

The summation over $j$ in Eq.~\ref{eq:coeff} does not begin at $j=1$. This is due to the property that the required cut-off order of the Chebyshev expansion increases with the time step $\tilde{\tau}$. As $N(\tilde{\tau})$ is the number of polynomial truncation order for a time step $\tilde{\tau}$, we introduce its inverse function $N^{-1}(n)$ to specify the maximum time time step for which the expansion is valid up to order $n$. Consequently, for a fixed $n$, any time step shorter than $\tilde{\tau} = N^{-1}(n)$ does not contribute to terms of order higher than $n$ in the summation of Eq.~\ref{eq:coeff}. The lower limit of the summation is therefore set to $j \ge N_0 = N^{-1}(n) / \tilde{\tau}$. This approach, which requires storing only the Chebyshev moments (complex numbers), reduces the additional memory cost to $O(N(\tilde{t}_{\mathrm{tot}}))$, which is negligible for memory usage.



For the cases where $\ket{\psi_1}\equiv\ket{\psi_2}$ in Eq.~\ref{eq:form2}, we can further exploit the property of the Chebyshev polynomials, 
\begin{equation}
    2T_m(\tilde{H})T_n(\tilde{H})=T_{m+n}(\tilde{H})+T_{|m-n|}(\tilde{H}), 
    \label{eq:cheby_prod}
\end{equation}
to further halve the number of expensive matrix-vector multiplications. The specific procedure, which computes higher-order moments from inner products of existing vectors, is given by
\begin{equation}
m_n = 
\begin{cases}
\langle \psi \ket{\phi_n}, & n \leq \frac{N(\tilde t_{\mathrm{tot}})}{2} \\
\braket{\phi_{\frac{n+1}{2}}}{\phi_{\frac{n-1}{2}}} - m_1, & n > \frac{N(\tilde t_{\mathrm{tot}})}{2},\ n\ \text{is odd} \\
\braket{\phi_{\frac{n}{2}}}{\phi_{\frac{n}{2}}} - m_0, & n > \frac{N(\tilde t_{\mathrm{tot}})}{2},\ n\ \text{is even}
\label{eq:cheby-moment}
\end{cases}
\end{equation}
where $\ket{\phi_n} = T_n(\tilde{H})\ket{\psi}$. This optimization is valid because $T_n(\tilde{H})$ is Hermitian, which implies $\bra{\phi_n} = \bra{\psi}T_n(\tilde{H})$.


\paragraph*{Energy-based Implementation:} The third strategy is designed for quantities requiring discrete energy sampling with the form of Eq.~\ref{eq:form1}.
In the sequential sPM, the calculation of $\ket{\Psi(E)}$ is formulated as follows:
\begin{equation}
\begin{split}
\ket{\Psi(E)} &=  \int_{-\infty}^{\infty}  a(E, t)\mathrm{e}^{-\mathrm{i}\hat H t}\ket{\psi_1} \, \mathrm{d}t \\
&\approx \tau \; a(E, 0) \ket{\psi_1}\\
&+\tau\sum_{j=1}^{N_t} \sum_{n=0}^{N(\tilde{\tau})} \big(a(E, t_j) c_n( \tilde{\tau}) T_n(\tilde{H})\ket{\psi_1(t_{j-1})}\\
&+a(E, -t_{j}) c_n(- \tilde{\tau}) T_n(\tilde{H})\ket{\psi_1(-t_{j-1})}\big).
\label{eq:form1_sequential}
\end{split}
\end{equation}
Similar as the case of state-based implementation,  the summation in the above expression can be reformulated by using the shared Chebyshev states
\begin{equation}
\ket{\Psi(E)} = 
\sum_{n=0}^{N(\tilde{t}_{\mathrm{tot}})} T_n(\tilde{H})\ket{\psi_1} C_n(E),
\label{eq:form2_concurrent}
\end{equation}
here the coefficient $C_n(E)$ is identical to that defined in the moment-based implementation (cf. Eq.~\ref{eq:coeff}) and consolidates information from both forward and backward propagation. 
In contrast to the moment-based method, this energy-based implementation operates on state sharing, making it particularly suitable for scenarios that require discrete sampling of the parameter $E$. 
A key advantage of this approach is its minimal memory overhead when the number of energy samples is small. 
Taking quasi-eigenstates as an example, more details will be discussed in Sec.~\ref{sec:QE}.

To validate the effectiveness of our new concurrent approach, we have implemented these three types of implementations for the calculations of several fundamental physical quantities within a tight-binding framework~\cite{2023-tbplas}, including the (local) density of states, quasi‐eigenstates, electronic conductivity, optical conductivity, dynamic polarization, and charge density. 

%% file: context/block_selection.tex
While the moment- and energy-based implementations circumvent the memory issue, the versatile state-based implementation still presents a trade-off between memory consumption and computational efficiency. We introduce time block to manage this trade-off systematically. This approach partitions the total propagation duration into uniform time blocks, with final-state propagation performed independently within each. The block size, $b$, thus becomes a tunable parameter to balance memory usage and acceleration.

The primary computational cost in sPM arises from algebraic operations on matrices and vectors. Assuming $\mathbf{M}$ is a sparse matrix, $\mathbf{x}$ and $\mathbf{y}$ are column vectors, and $a$ is a scalar, we can categorize these costs as follows.
The first category is the matrix-vector multiplication, $\mathbf{y} \leftarrow a\mathbf{M}\mathbf{x} - \mathbf{y}$, denoted as $\texttt{amxsy}$.
The second is the vector scaling and addition, $\mathbf{y} \leftarrow a\mathbf{x} + \mathbf{y}$, denoted as $\texttt{axpy}$.
The third is the inner product between vectors, denoted as $\texttt{dot}$.
Final-state propagation significantly reduces the number of computationally intensive $\texttt{amxsy}$ operations by introducing a certain number of $\texttt{axpy}$ and $\texttt{dot}$ operations, leading to a substantial improvement in efficiency. 

To theoretically determine the optimal block size $b_{\mathrm{opt}}$, we estimate the computational cost of the propagation part by analyzing the number of main mathematical operations and their average cost of single call. 
The former depends on the selection of block size $b$ and the number of time steps $N_t$, while the latter is closely related to the matrix density $D_H$, defined as the average number of non-zero elements per row of the Hamiltonian.
This approach allows us to predict the computational cost, $C(b,N_t,D_H)$, and thereby identify the optimal $b$ that minimizes this cost.


%
The computational cost function for the time-propagation part, $C(b,N_t,D_H)$, is defined as:
\begin{equation}
\begin{aligned}
    C(b,N_t,D_H) &= N_{\texttt{amxsy}}(b,N_t) \cdot t_{\texttt{amxsy}}(D_H) \\
    &+ N_{\texttt{axpy}}(b,N_t) \cdot t_{\texttt{axpy}}(D_H) + C_{\texttt{other}}(N_t,D_H).
\end{aligned}
\end{equation}
Here, $N_{\texttt{amxsy}}(b,N_t)$, $N_{\texttt{axpy}}(b,N_t)$ denote the total number of corresponding algebraic operation. Since $N_{\texttt{dot}}(N_t)$ is independent of the block size $b$, its contribution appears in a constant offset $C_{\texttt{other}}(N_t,D_H)$. The corresponding average cost $t_{\texttt{amxsy}}(D_H)$ and $t_{\texttt{axpy}}(D_H)$ can be obtained via benchmarking prior to the main calculation for a given $D_H$. 
As demonstrated in Tab.~\ref{tab:cost}, $N_{\texttt{amxsy}}(b,N_t)$, $N_{\texttt{axpy}}(b,N_t)$ and $N_{\texttt{dot}}(N_t)$ can be analytically precomputed prior to actual calculations for different physical quantities.
The optimal block size, $b_{\mathrm{opt}}$, is then chosen to minimize this cost function for given $N_t$ and $D_H$:
\begin{equation}
    b_{\mathrm{opt}} = \arg\min_{b} C(b,N_t,D_H).
\end{equation}
It is important to note that increasing the block size also increases memory consumption. Owing to the rapid decrease of the expansion load for small time step, one can achieve substantial efficiency gains at a modest memory cost.

It should be noted that $b_{\mathrm{opt}}$ cannot achieve an overall computational speedup ratio of $n=C(1,N_t,D_H)/C(b_{\mathrm{opt}},N_t,D_H)$, since there exists a non-accelerable component in the computation that cannot be optimized by the propagation strategy.
This scenario is analogous to the limits on parallel computing, which can be modeled by Amdahl's law~\cite{1967-amdahl}:
\begin{equation}
    S = \frac{1}{1 - P + \frac{P}{n}}.
\label{eq:amdahl}
\end{equation}
In this model, $n$ represents the theoretical speedup of the accelerable portion (i.e., time propagation). The parameter $P$ is the fraction of the total computation time that this accelerable portion represents within the baseline sequential sPM. Unlike $n$, which is calculated theoretically, $P$ must be determined empirically by fitting the model to the measured overall speedup $S(n)$. Crucially, the value of $P$ affects the achievable overall speedup $S$ but does not alter the selection of the optimal block size $b_{\mathrm{opt}}$, as the latter is determined solely by minimizing the cost function of the time-propagation part. In Sec.~\ref{sec:QE}--\ref{sec:CD}, we will present benchmark results for the achieved speedup and use this model to fit the parameter $P$ for the various physical quantities. In addition to efficiency, validation benchmark will also be performed.

All calculations are performed using a customized version of the TBPLaS software package~\cite{2023-tbplas}, where time propagations are implemented using Chebyshev polynomial expansion.
The primary test system is a $(20 \times 20)$ supercell of a magic-angle twisted bilayer graphene system~\cite{tbg} with a $1.05^\circ$ twist angle, comprising 4,763,200 atoms within a $p_z$ orbital model. This system will be used for all physical quantities discussed in this work unless otherwise noted. 
To reduce statistical errors from single runs, all timing and memory benchmarks are averaged over multiple runs with multiple random states. Specifically, for benchmarks against the number of time steps, $N_t$, we use $4096/N_t$ states (for $N_t<4096$), while for tests against matrix density, $D_H$, we use 10, 5, and 2 states for $D_H=60, 120,$ and $230$, respectively. For tests on other parameters discussed in subsequent sections, such as the block size $b$ and the number of quasi-eigenstates $N_E$, 5 random states are used. This testing protocol is applied consistently throughout the paper.
All benchmarks were executed on a single cluster node equipped with two Intel\textregistered{} Xeon\textregistered{} Gold 6548Y+ processors (64 cores total) and 256 GB of RAM.

%% file: context/DOS.tex
In the sPM, the density of states (DOS) is given by~\cite{2000-hams-eigenvalues,2010-yuan-graphene}
\begin{equation}
    D(E) = \frac{1}{2\pi}\int_{-\infty}^{\infty}\mathrm{e}^{\mathrm{i}Et}C^{\mathrm{DOS}}(t) \,\mathrm{d}t,
\end{equation}
where $C^{\mathrm{DOS}}(t)=\braket{\psi}{\psi(t)}$ is the DOS correlation function for a random state $\ket{\psi}$. The spectral width of the DOS matches that of the Hamiltonian $H$, which sets the rescaled time step to $\tilde{\tau}_{\mathrm{DOS}}=\pi$.  The total propagation duration $t_{\mathrm{tot}}=N_{t}\cdot\tau_{\mathrm{DOS}}$ determines the energy resolution of the DOS, and a window function is typically employed to mitigate truncation artifacts from the finite time cutoff. Although sample averaging over multiple random states is typically necessary to mitigate stochastic errors in finite systems, the present work concentrates on optimizing single-sample propagation. Consequently, we deliberately omit the sample averaging notation throughout the paper.

Since $C^{\mathrm{DOS}}(t)$ is the most time-consuming component of the DOS calculation, once $C^{\mathrm{DOS}}(t)$ is obtained, the DOS can be rapidly computed via the inverse Fourier transform. Exploiting the property $C^{\mathrm{DOS}}(-t)={C^{\mathrm{DOS}}}^*(t)$, values of $C^{\mathrm{DOS}}(t)$ for $t<0$ are obtained as the complex conjugates of those for $t>0$. The conventional approach, which we term the DOS-Sequential method, computes the correlation function $C^{\mathrm{DOS}}(t)$ by sequentially propagating the state $\ket{\psi(t)}$ at each time step.


In contrast, the new approach utilizes the moment-based implementation—which we term the DOS-Concurrent-M method—as the underlying correlation function has the applicable form $\bra{\psi_1}\mathrm{e}^{-\mathrm{i}\Hamil t}\ket{\psi_2}$. This implementation circumvents the explicit calculation of intermediate states $\ket{\psi(t)}$ by instead computing the $N(\tilde{t}_{\mathrm{tot}})$ Chebyshev moments $m_n = \braket{\psi}{T_n(\tilde{H})|\psi}$. The correlation function at any earlier time $t_j$ ($j \cdot \tau_{\mathrm{DOS}}$) can then be rapidly reconstructed as a linear combination of these pre-computed moments:
\begin{equation}
\begin{split}
    C^{\mathrm{DOS}}(t_j) &= \braket{\psi}{\psi(t_j)} \\
    &= \bra{\psi}\sum_{n=0}^{N(\tilde{t}_{j})} c_n(\tilde{t}_j) T_n(\tilde{H})\ket{\psi} 
    = \sum_{n=0}^{N(\tilde{t}_j)} c_n(\tilde{t}_j) m_n.
\end{split}
\end{equation}

This approach requires storing only $N(\tilde{t}_{\mathrm{tot}})$ real numbers, rendering the memory overhead negligible. It therefore allows the entire propagation to be treated as a single block ($b=N_t$), achieving a theoretical speedup of approximately $R(\pi)/R(\infty)\approx 6$ over the sequential method.

Moreover, the combination of the moment-based implementation ($\sim$6$\times$ speedup) and the cost-halving optimization for symmetric correlation functions (Eq.~\ref{eq:cheby-moment}) allows the new method to achieve a total theoretical speedup of approximately 12$\times$ relative to the sequential sPM, without incurring additional memory overhead. 
Furthermore, we note that the calculation of the local density of states (LDOS)~\cite{2020-shi-nc} is analogous; for the LDOS at a position $\mathbf{r}$, one simply replaces the random state $\ket{\psi}$ with a localized state $\ket{\mathbf{r}}$. We therefore do not discuss the LDOS implementation separately.

\begin{figure}[htbp]
    \centering
    \includegraphics[width=0.8\columnwidth]{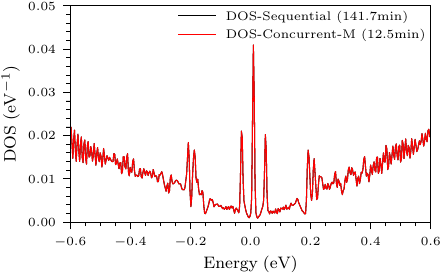}
    \caption{
    \textbf{Accuracy validation of the new concurrent method for the density of states calculation.} 
    The density of states (DOS) from the DOS-Concurrent-M method (red line) is compared against the baseline DOS-Sequential method (black line). The calculation was performed on a magic-angle twisted bilayer graphene supercell containing 4,763,200 atoms ($D_H=230$, $N_t=4096$). The wall time for each method is provided in parentheses within the legend.
    }
    \label{fig:acc_dos}
\end{figure}

\begin{figure*}[htbp]
    \centering
    \begin{overpic}[width=0.47\textwidth]{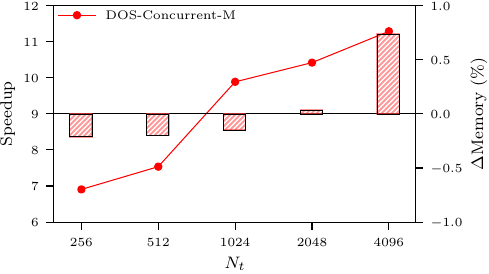}
        \put(-2,52.7){\bfseries (a)}
    \end{overpic}
    \quad
    \begin{overpic}[width=0.45\textwidth]{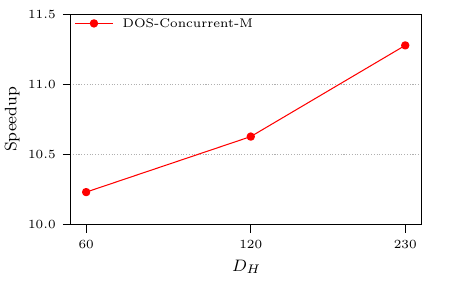}
        \put(-2,55){\bfseries (b)}
    \end{overpic}
    
    \caption{%
    \textbf{Performance comparison of the DOS-Concurrent-M method vs. the baseline DOS-Sequential method.}
    (a) Speedup (line) and relative memory consumption (bars) as a function of the total number of time steps $N_t$ ($D_H=230$).
    (b) Speedup as a function of the Hamiltonian matrix density $D_H$ ($N_t=4096$).
    }
    \label{fig:DOS_combined}
\end{figure*}

First, we validate the method's accuracy for the DOS calculation. As confirmed in Fig.~\ref{fig:acc_dos}, the DOS calculated by the DOS-Concurrent-M method is in excellent agreement with the baseline result. 
To quantitatively assess accuracy, we use the infinity norm ($\varepsilon_{\infty}$) to measure the maximum absolute error against the baseline results: 
\begin{equation}
\varepsilon_{\infty}(A,B)
\;=\;
\bigl\|A - B\bigr\|_{\infty}
\;=\;
\max_{i}\bigl|A_{i} - B_{i}\bigr|,
\end{equation}
where $A$ and $B$ represent the results from the new and baseline methods, respectively.
The infinity norm reveals an error of only $8.3 \times 10^{-14}$ for correlation function $C^{\mathrm{DOS}}(t)$ and $2.8 \times 10^{-13}$ for the DOS $D(E)$. These error values approach machine precision, confirming that the new method introduces no discernible loss of accuracy.

%
Having established its accuracy, we now evaluate its computational performance and memory costs. Owing the moment-based implementation, the new method incurs no additional memory consumption for calculating the DOS. Consequently, time block is unnecessary, meaning the block size $b$ can be set equal to the total number of time steps, $N_t$. As shown by the bar chart in Fig.~\ref{fig:DOS_combined}(a), the memory overhead of the DOS-Concurrent-M method compared to the baseline DOS-Sequential method is less than 1\%. For shorter propagation durations, it can even exhibit a slight reduction in memory usage. Meanwhile, the speedup of  DOS-Concurrent-M over the DOS-Sequential method increases with the total propagation duration, reaching over 11$\times$ at $N_t=4096$, which approaches the theoretical limit of a 12$\times$ speedup.
The tests described above were based on a Hamiltonian with $D_H=230$. Fig.~\ref{fig:DOS_combined}(b) further illustrates the dependence of the speedup on $D_H$. For the same total propagation duration of $N_t=4096$, a denser Hamiltonian matrix results in a more speedup. Specifically, the speedup increases from 10.2$\times$ (for $D_H=60$) to 11.3$\times$ (for $D_H=230$).
These tests demonstrate that the new method holds a significant advantage for DOS calculations. Compared to the baseline sequential sPM, it can achieve a speedup of over 10$\times$ without additional memory cost or any loss of precision.

%% file: context/QE.tex
Quasi‐eigenstates (QE) $\ket{\Psi(E)}$ are approximations to linear combinations of all eigenstates with energy $E$, used to analyze the real‐space distribution of electronic states at different energies and to simulate scanning tunneling spectroscopy experiments~\cite{2020-shi-nc}. In the sPM, the quasi‐eigenstates $\ket{\Psi(E)}$ is computed as~\cite{2010-yuan-graphene}:
\begin{equation}
\ket{\Psi(E)} = \frac{1}{2\pi} \int_{-\infty}^{\infty} \mathrm{e}^{\mathrm{i}Et} |\psi(t)\rangle \, \mathrm{d}t.
\label{eq:QE}
\end{equation}

Similar as the calculation of DOS, the spectral width of the quasi‐eigenstates matches that of the Hamiltonian $H$, setting the rescaled time step to $\tilde{\tau}_{\mathrm{QE}} = \pi$. The total propagation duration $t_{\mathrm{tot}}$ must be sufficiently long to resolve the desired energy $E$; otherwise, $\ket{\Psi(E)}$ will contain contributions from eigenstates at neighboring energies. The sequential sPM, which we term the QE-Sequential method, computes this integral by sequentially propagating the state $\ket{\psi}$ forward and backward in time.

The new method offers two distinct implementations for QE calculations. The first is the general-purpose state-based implementation (QE-Concurrent-S). In this approach, we propagate directly to the final states $\ket{\psi( \pm t_{\mathrm{tot}})}$ and then reconstruct any intermediate state $\ket{\psi(\pm t_{j})}$ as a linear combination of the shared Chebyshev states. This incurs a memory overhead of $O(N_{\mathrm{basis}}\cdot N_t)$, which can become a bottleneck in large-scale simulations, making an optimized time-blocking strategy critically important. 

However, for calculating QE, which often require sampling at only a few discrete energies, the intermediate states do not need to be explicitly reconstructed. This allows for a more efficient energy-based implementation (QE-Concurrent-E). Starting from the final-state Chebyshev states, this method incorporates the Fourier transform from Eq.~\ref{eq:QE} directly into the reconstruction coefficients, yielding energy-dependent modulated Bessel coefficients $C_n^{\mathrm{QE}}(E)$ (derived from Eq.~\ref{eq:coeff}):
\begin{equation}
    C_n^{\mathrm{QE}}(E) = \tau \sum_{j=N_0}^{N_t} \frac{(2-\delta_{j,0})}{2} \big( \mathrm{e}^{\mathrm{i}Et_j} c_n( \tilde{t}_j) 
     + \mathrm{e}^{-\mathrm{i}Et_j} c_n( -\tilde{t}_{j}) \big).
\label{eq:QE_coeff}
\end{equation}
The quasi‐eigenstates at any target energy can then be computed directly from the Chebyshev states: 
\begin{equation}
\ket{\Psi(E)} = \sum_{n=0}^{N(\tilde{t}_{\mathrm{tot}})} C_n^{\mathrm{QE}}(E)\,T_n(\tilde{H})\ket{\psi}.
\end{equation}
This procedure
avoids the explicit storage of intermediate states.

Although both implementations leverage the new concurrent propagation strategy, they exhibit different strengths.
The QE-Concurrent-S method requires an additional memory overhead of $O(N_{\mathrm{basis}}\cdot b)$, where $b$ is the block size; consequently, it cannot employ arbitrarily large blocks, limiting its speedup to below 6$\times$. In contrast, QE-Concurrent-E requires no additional memory, allowing the block size $b$ to equal the total number of time steps $N_t$ and achieve a 6$\times$ speedup. Moreover, QE-Concurrent-E merges the forward and backward time contributions within the coefficients of Eq.~\ref{eq:QE_coeff}, which further halves the computational cost, leading to a total theoretical speedup of approximately 12$\times$. However, this advantage diminishes when a large number ($N_E$) of quasi‐eigenstates must be calculated. The computational cost of QE-Concurrent-E scales as $O(N(\tilde{t}_{\mathrm{tot}})\cdot N_E)$, whereas QE-Concurrent-S scales as $O(N_t\cdot N_E)$. Since the number of expansion terms $N(\tilde{t}_{\mathrm{tot}})$ is significantly larger than the number of time steps $N_t$ (as $N(\tilde{t}_{\mathrm{tot}})/N_t \approx R(\tilde{t}_{\mathrm{tot}})\cdot\pi > \pi$), the larger prefactor causes the efficiency advantage of QE-Concurrent-E to decrease as $N_E$ increases.


\begin{figure}[htbp]
    \centering
    \includegraphics[width=0.8\columnwidth]{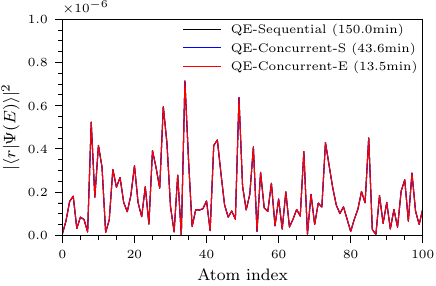}
    \caption{
    \textbf{Accuracy validation of the new concurrent method for the quasi-eigenstates calculation.} 
    The quasi-eigenstates (QE) calculated by the QE-Concurrent-S (blue line) and QE-Concurrent-E (red line) methods are compared against the baseline QE-Sequential method (black line). For visual clarity, the plot displays the results for a single quasi-eigenstates on the first 100 atoms. The calculation was performed on a magic-angle twisted bilayer graphene supercell containing 4,763,200 atoms ($D_H=120$, $N_t=4096$, $N_E=5$). The wall time for each method is provided in parentheses within the legend.}
\label{fig:acc_qe}
\end{figure}

Fig.~\ref{fig:acc_qe} compares the QE-Concurrent-S and QE-Concurrent-E results with the baseline QE-Sequential calculation, showing excellent agreement across all methods. The consistency is further quantified by the infinity norm ($\varepsilon_{\infty}$), which remains below $1.3 \times 10^{-17}$ for both QE-Concurrent-S and QE-Concurrent-E relative to the baseline, confirming their high numerical accuracy.

\begin{figure*}[htbp]
    \centering
    \begin{overpic}[width=0.47\textwidth]{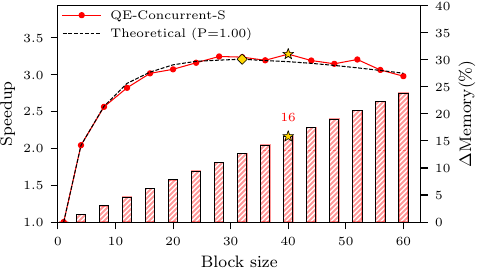}
        \put(-2,52.7){\bfseries(a)}
    \end{overpic}
    \quad
    \begin{overpic}[width=0.47\textwidth]{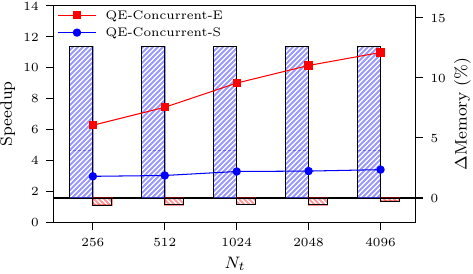}
        \put(-2,52.7){\bfseries(b)}
    \end{overpic}

    \vspace{0ex}  

    \begin{overpic}[width=0.48\textwidth]{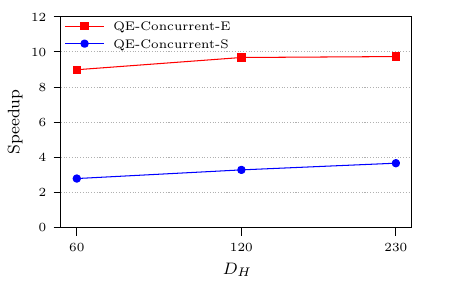}
        \put(0,52.7){\bfseries(c)}
    \end{overpic}
    \quad
    \begin{overpic}[width=0.47\textwidth]{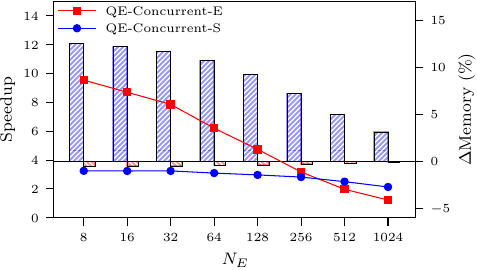}
        \put(-2,52.7){\bfseries(d)}
    \end{overpic}

  \caption{%
  \textbf{Performance comparison of the QE-Concurrent-S/QE-Concurrent-E methods vs. the baseline QE-Sequential method.}
    (a) Speedup (line) and relative memory consumption (bars) of the QE-Concurrent-S method as a function of block size $b$. Asterisks and diamonds mark the empirically measured and theoretical optimal values, respectively ($D_H=120$, $N_t=1024$, $N_E=5$).
    (b) Speedup (lines) and relative memory consumption (bars) of the QE-Concurrent-S and QE-Concurrent-E methods at the optimal block size $b$, as a function of the total number of time steps $N_t$ ($D_H=120$, $N_E=5$).
    (c) Speedup of the QE-Concurrent-S and QE-Concurrent-E methods at the optimal block size $b$, as a function of the Hamiltonian matrix density $D_H$ ($N_t=1024$, $N_E=5$).
    (d) Speedup (lines) and relative memory consumption (bars) of the QE-Concurrent-S and QE-Concurrent-E methods at the optimal block size $b$, as a function of the number of quasi-eigenstates $N_E$ ($D_H=120$, $N_t=1024$).
  }
    \label{fig:QE_combined}
\end{figure*}

Fig.~\ref{fig:QE_combined} presents a comparison between the two proposed implementations and the baseline QE-Sequential method.
Due to its additional memory consumption, the QE-Concurrent-S method requires a time-blocking strategy. We first discuss the selection of the optimal time block size $b$. Fig.~\ref{fig:QE_combined}(a) shows the speedup and memory consumption of QE-Concurrent-S as a function of $b$, for a total of $N_t=1024$ steps. According to our tests, the speedup of QE-Concurrent-S exhibits a peak behavior, first increasing and then decreasing, which is consistent with the theoretical curve fitted based on Amdahl's law. The optimal efficiency is empirically measured at $b=40$, achieving a speedup of approximately 3.3$\times$ compared to the baseline QE-Sequential method. This is very close to the 3.2$\times$ speedup at the theoretical optimum of $b=32$. This optimal block size introduces a memory overhead of only about 16\%.

Using the optimal $b$, Fig.~\ref{fig:QE_combined}(b) tests the performance of the new methods for various $N_t$, with the QE-Concurrent-E method also included in the comparison. As shown, the speedup for both new methods increases with $N_t$, though the performance of QE-Concurrent-E is more sensitive to this parameter. Furthermore, the memory consumption for both methods remains nearly constant for different $N_t$: QE-Concurrent-S incurs a small overhead of approximately 10\%, while QE-Concurrent-E introduces no additional memory cost.

We also investigated the dependence of the speedup on $D_H$ (Fig.~\ref{fig:QE_combined}(c)). Similar to the DOS calculations, the speedups for both QE-Concurrent-S and QE-Concurrent-E improve as $D_H$ increases. Specifically, the speedup for QE-Concurrent-S increases from 2.8$\times$ to 3.7$\times$, while for QE-Concurrent-E it increases from 9.0$\times$ to 9.7$\times$. This trend aligns with our theoretical analysis: increasing the $D_H$ of the Hamiltonian increases the time ratio of $\texttt{amxsy}$ to $\texttt{axpy}$ operations, thereby improving the overall speedup and causing it to approach the theoretical limit.

%
The above tests considered cases with a small number of quasi-eigenstates ($N_E=5$). Fig.~\ref{fig:QE_combined}(d) examines the speedup and relative memory consumption of QE-Concurrent-S and QE-Concurrent-E as the number of quasi-eigenstates ($N_E$) increases. As the figure shows, the speedup for both methods decreases as $N_E$ increases, eventually falling to 2.1$\times$ (QE-Concurrent-S) and 1.2$\times$ (QE-Concurrent-E) at $N_E=1024$.
This decrease occurs because the number of $\texttt{axpy}$ operations grows rapidly with $N_E$ while the number of $\texttt{amxsy}$ operations remains constant, thus increasing the computational share of the non-propagation part. In the context of Amdahl's law, this corresponds to a decrease in the accelerable fraction $P$. 
The speedup of QE-Concurrent-E decreases more rapidly because its complexity scales as $O(N_E \cdot N(\tilde{t}_{\mathrm{tot}}))$, based on the Chebyshev states. In contrast, the QE-Sequential and QE-Concurrent-S methods scale as $O(N_E \cdot N_t)$, based on the propagated states. As the number of expansion terms $N(\tilde{t}_{\mathrm{tot}})$ is much larger than the number of time steps $N_t$, the number of $\texttt{axpy}$ operations in QE-Concurrent-E grows more quickly with $N_E$, leading to a faster reduction in the speedup.
Regarding memory differences, the relative overhead of QE-Concurrent-S decreases from 12.5\% to 3.1\%. This is because the additional memory required by QE-Concurrent-S (compared to QE-Sequential) is independent of $N_E$. As $N_E$ increases, the total memory usage grows, thus reducing the relative percentage of this fixed overhead. The memory usage of QE-Concurrent-E, in contrast, remains almost unchanged, shifting from -0.5\% to -0.1\%, which is still negligible.

In summary, the new method provides two powerful implementations for QE computation. QE-Concurrent-E is the optimal choice for a small number of quasi-eigenstates ($N_E < 200$), where it can achieve a speedup approaching 10$\times$ without requiring additional memory. However, for $N_E > 300$, the state-based QE-Concurrent-S method begins to hold the advantage in efficiency. At its optimal block size, it provides a speedup of approximately 3$\times$ while requiring only a small amount of extra memory ($\sim 10\%$).

%% file: context/DC.tex
Electronic conductivity (EC) is an important measure of a material's electrical transport properties and can be obtained from the Chester--Thellung formula~\cite{1959-Chester-Thellung}. At zero temperature, the diagonal component of the electronic conductivity in direction $\alpha$ is given by~\cite{2010-yuan-graphene}
\begin{equation}
\begin{split}
\sigma_{\alpha\alpha}(E) &= \lim_{\tau \to \infty} \sigma_{\alpha\alpha}(E, \tau) \\
&= \lim_{\tau \to \infty} \frac{D(E)}{A} \int_0^\tau \mathrm{Re} \left[ \mathrm{e}^{-\mathrm{i}Et} C^{\mathrm{EC}}_{\alpha}(t,E) \right] \, \mathrm{d}t,
\end{split}
\label{eq:DC}
\end{equation}
where $D(E)$ is the density of states, $A$ is the area (2D) or volume (3D) of the system. The EC correlation function is defined as
\begin{equation}
C^{\mathrm{EC}}_{\alpha}(t,E) = \frac{\langle \psi| \hat J_\alpha \mathrm{e}^{\mathrm{i}\hat{H}t} \hat J_\alpha | \Psi(E) \rangle}{| \langle \psi | \Psi(E) \rangle |},
\label{eq:dc_corr}
\end{equation}
where $\ket{\Psi(E)}$ is the quasi‐eigenstates obtained from the random state $\ket{\psi}$. The current density operator $\hat{J}_\alpha$ is defined via the commutator $\hat{J}_\alpha = (\mathrm{i}/\hbar)[\hat{H},\hat{P}_\alpha]$, where the polarization operator is $\hat{P}_\alpha = -e \sum_{i=1}^{N_{\mathrm{basis}}} \hat{r}_{\alpha,i}$, with $\hat{r}_{\alpha,i}$ being the position operator of the $i$-th electron in direction $\alpha$. 

In a practical calculation, the quasi‐eigenstates $\ket{\Psi(E)}$ depends on the energy $E$, and multiple quasi‐eigenstates are often required. To minimize the computational cost, time-evolution operator is applied only to the bra state in Eq.~\ref{eq:dc_corr}. This requires the calculation of two vectors:
\begin{align}
\ket{\psi_{1,\alpha}^{\mathrm{EC}}(t)}&=\mathrm{e}^{-\mathrm{i}\hat{H}t}\hat J_\alpha\ket{\psi},\\
\ket{\psi_{2,\alpha}^{\mathrm{EC}}(E)}&=\hat J_\alpha\ket{\Psi(E)}.
\end{align}

Since $E$ can be any value within the spectrum of $\hat{H}$, the rescaled time step is set to $\tilde{\tau}_{\mathrm{EC}}=\pi$. The sequential sPM, which we term the EC-Sequential method, computes the correlation function by sequentially propagating $\ket{\psi_{1,\alpha}^{\mathrm{EC}}(t)}$ at each time step.

The new method offers two more efficient implementations. The first is the general-purpose state-based implementation (EC-Concurrent-S).
Because the EC correlation function has the requisite form $\bra{\psi_1}\mathrm{e}^{-\mathrm{i}\hat{H}t}\ket{\psi_2}$, the moment-based implementation (EC-Concurrent-M) is also applicable. This approach avoids the reconstruction of intermediate states by directly computing the Chebyshev moments from the final state:
\begin{equation}
    m^{\mathrm{EC}}_{n,\alpha}(E) = \bra{\psi_{1,\alpha}^{\mathrm{EC}}}T_n(\tilde{H})\ket{\psi_{2,\alpha}^{\mathrm{EC}}(E)}.
    \label{eq:dc_mom}
\end{equation}

Since the bra and ket vectors in this definition differ, the cost-halving property of Eq.~\ref{eq:cheby_prod} cannot be exploited here. The correlation function is then reconstructed from the moments:
\begin{equation}
    C^{\mathrm{EC}}_{\alpha}(t,E) = \frac{1}{|\langle \psi|\Psi(E)\rangle|} \sum_{n=0}^{N(\tilde t)} c_n^*(\tilde t)\,m^{\mathrm{EC}}_{n,\alpha}(E).
\end{equation}

The EC-Concurrent-M method significantly reduces memory usage compared to EC-Concurrent-S. However, similar to QE-E, its computational cost scales with the number of expansion terms, $N(\tilde{t}_{\mathrm{tot}})$, which causes its efficiency advantage over EC-Concurrent-S (whose cost scales with $N_t$) to diminish as the number of quasi‐eigenstates ($N_E$) grows.

\begin{figure}[htbp]
    \centering
    \includegraphics[width=0.8\columnwidth]{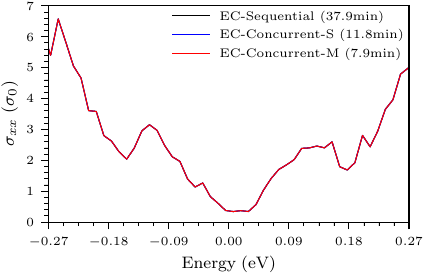}
    \caption{
    \textbf{Accuracy validation of the new concurrent method for the electronic conductivity calculation.} 
    The electronic conductivity (EC) calculated by the EC-Concurrent-S (blue line) and EC-Concurrent-M (red line) methods are compared against the baseline EC-Sequential method (black line). The calculation was performed on a magic-angle twisted bilayer graphene supercell containing 4,763,200 atoms ($D_H=120$, $N_t=1024$, $N_E=49$). The wall time for each method is provided in parentheses within the legend. The electronic conductivity is in units of $\sigma_0=e^2/\hbar$.
}
\label{fig:acc_dc}
\end{figure}

\begin{figure*}[htbp]
    \centering
    \begin{overpic}[width=0.47\textwidth]{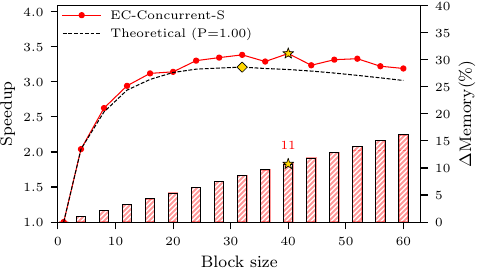}
        \put(-2,52.7){\bfseries(a)}
    \end{overpic}
    \quad
    \begin{overpic}[width=0.46\textwidth]{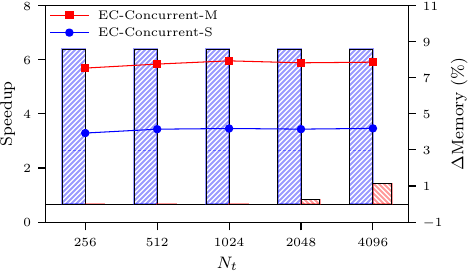}
        \put(-2,53.6){\bfseries(b)}
    \end{overpic}

    \vspace{0ex}  

    \begin{overpic}[width=0.47\textwidth]{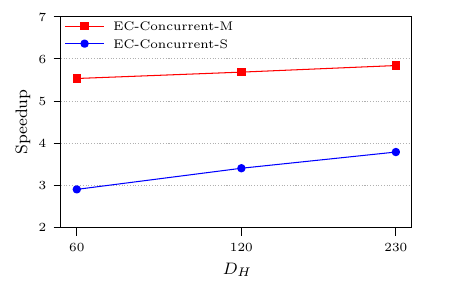}
        \put(0,52.7){\bfseries(c)}
    \end{overpic}
    \quad
    \begin{overpic}[width=0.46\textwidth]{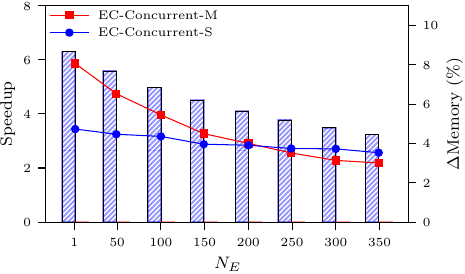}
        \put(-2,53.6){\bfseries(d)}
    \end{overpic}

  \caption{%
  \textbf{Performance comparison of the EC-Concurrent-S/EC-Concurrent-M methods vs. the baseline EC-Sequential method.}
    (a) Speedup (line) and relative memory consumption (bars) as a function of block size $b$. Asterisks and diamonds mark the empirically measured and theoretical optimal values, respectively ($D_H=120$, $N_t=1024$, $N_E=5$).
    (b) Speedup (lines) and relative memory consumption (bars) at the optimal block size $b$, as a function of the total number of time steps $N_t$ ($D_H=120$, $N_E=5$).
    (c) Speedup at the optimal block size $b$, as a function of the Hamiltonian matrix density $D_H$ ($N_t=1024$, $N_E=5$).
    (d) Speedup (lines) and relative memory consumption (bars) at the optimal block size $b$, as a function of the number of quasi-eigenstates $N_E$ ($D_H=120$, $N_t=1024$).
  }
    \label{fig:DC_combined}
\end{figure*}

For the electronic conductivity calculation, we tested the state-based implementation (EC-Concurrent-S) and the moment-based implementation (EC-Concurrent-M). 
To validate the accuracy of the EC approaches, Fig.~\ref{fig:acc_dc} displays the close agreement between the EC-Concurrent-S and EC-Concurrent-M results and the baseline. Quantitative error assessment using the infinity norm supports this observation, with $\varepsilon_{\infty}$ below $2.5 \times 10^{-13}$ for the correlation function $C^{\mathrm{EC}}_{\alpha}(t,E)$ and below $1.2 \times 10^{-11}$ for the electronic conductivity $\sigma_{\alpha\alpha}(E)$. 
%

Fig.~\ref{fig:DC_combined} compares their performance against the baseline EC-Sequential method, showing the computational and memory costs for various values of $b$, $N_t$, $D_H$, and $N_E$.
As shown in Fig.~\ref{fig:DC_combined}(a), similar to QE-Concurrent-S, the speedup of EC-Concurrent-S also peaks at $b=40$, achieving a speedup of approximately 3.5$\times$ with a memory overhead of only 11\%. This differs by just 0.02 from the speedup at the theoretical optimum of $b=32$ predicted by the fitted curve. We note a certain discrepancy between the measured performance of the EC-Concurrent-S method and the theoretical curve fitted using Amdahl's law, for which there are two main reasons. First, the time cost for the EC calculation does not include the time required to obtain the quasi-eigenstates or the density of states. Its initialization only requires a single application of the current density operator per direction, which is very fast. In contrast, the initialization for other quantities like dynamical polarization requires the application of the time-consuming Fermi-Dirac operator expanded in Chebyshev polynomials. Second, the introduction of time blocks allows operations such as inner products and current density operator applications—which are dispersed in the EC-Sequential method—to be consolidated within a single loop. This improves cache reuse and the parallel efficiency of these operations, an effect not accounted for in the theoretical speedup estimate for the propagation part. Therefore, when the block size $b$ is large, the empirically observed speedup can exceed the theoretical prediction.

%
%
%
Fig.~\ref{fig:DC_combined}(b) shows that the speedups for both new methods increase as the total propagation duration increases, with the EC-Concurrent-M speedup approaching its theoretical limit of 6$\times$. The relative memory overhead of EC-Concurrent-S remains nearly constant at $\sim 10\%$ across different $N_t$, while EC-Concurrent-M introduces no extra overhead for $N_t \le 1024$ and only a negligible $\sim 1\%$ overhead at $N_t=4096$.
Fig.~\ref{fig:DC_combined}(c) shows that the speedups for both new methods improve as the matrix density $D_H$ increases.
For a small number of quasi-eigenstates ($N_E=5$), the speedup of EC-Concurrent-M is significantly better than that of EC-Concurrent-S. This advantage diminishes as $N_E$ increases, with a crossover in efficiency occurring around $N_E \approx 200$ (see Fig.~\ref{fig:DC_combined}(d)).

We therefore recommend prioritizing the EC-Concurrent-M method for the EC calculations, with EC-Concurrent-S serving as an alternative for applications requiring quasi-continuous sampling.

%% file: context/OC.tex
Optical conductivity (OC) is a key quantity characterizing charge transport under an electromagnetic field. According to the Kubo formula~\cite{1957-kubo}, the real part of the OC tensor in direction $\alpha$ due to a field in direction $\beta$ (excluding the Drude contribution at $\omega=0$) is given by~\cite{2010-yuan-graphene}
\begin{equation}
\begin{split}
\mathrm{Re}\,\sigma_{\alpha\beta}(\hbar\omega) 
&= \lim_{E \to 0^+} 
\frac{2\bigl(\mathrm{e}^{-\hbar\omega/k_B T} - 1\bigr)}{\hbar \omega \,A} \\
&\quad \times \int_0^\infty \mathrm{e}^{-Et} \sin(\omega t)\,\mathrm{Im}\,C^{\mathrm{OC}}_{\alpha\beta}(t)\,\mathrm{d}t,
\end{split}
\end{equation}
where $T$ is the temperature and $A$ is the area (2D) or volume (3D) of the system, and the imaginary part can be obtained with the Kramers--Kronig relation
\begin{equation}
\mathrm{Im}\,\sigma_{\alpha\beta}(\hbar\omega)
= -\frac{1}{\pi}\,\mathcal{P}
\int_{-\infty}^{\infty}
\frac{\mathrm{Re}\,\sigma_{\alpha\beta}(\hbar\omega')}{\omega' - \omega}
\,\mathrm{d}\omega'.
\end{equation}

The core of this calculation is the OC correlation function, defined as
\begin{equation}
    C^{\mathrm{OC}}_{\alpha\beta}(t) = \bigl\langle \psi_1^{\mathrm{OC}}(t)\,\bigr|\hat J_\alpha\bigl|\psi_{2,\beta}^{\mathrm{OC}}(t)\bigr\rangle,
\label{eq:oc_corr}
\end{equation}
which requires the independent time propagation of two states constructed from a random state 
\begin{align}
    \ket{\psi_1^{\mathrm{OC}}(t)} &= \mathrm{e}^{-\mathrm{i}\hat H t}\,f(\hat H)\,\ket{\psi},\\
    \ket{\psi_{2,\beta}^{\mathrm{OC}}(t)} &= \mathrm{e}^{-\mathrm{i}\hat H t}\,\bigl[1 - f(\hat H)\bigr]\,\hat J_\beta\ket{\psi},
\end{align}
where $f(\hat{H})$ and $\bigl[1 - f(\hat H)\bigr]$ are the Fermi-Dirac operators.

Theoretically, the spectral width of the OC corresponds to the energy differences between occupied and unoccupied states. However, to avoid numerical artifacts such as aliasing, we conservatively set the spectral width to twice that of the Hamiltonian. This results in a time step of $\tau_{\mathrm{OC}}=\pi/(2|H|)$, which corresponds to a rescaled time step of $\tilde{\tau}_{\mathrm{OC}}=\pi/2$. The conventional sequential sPM, which we term the OC-Sequential method, computes the correlation function by propagating both states sequentially.

Due to the form of OC correlation function, the OC calculation can only be accelerated using our general-purpose state-based implementation (OC-Concurrent-S).
Since the rescaled time step $\tilde{\tau}_{\mathrm{OC}} = \pi/2 < \pi$, the theoretical speedup for the time-propagation portion of the calculation is expected to exceed 6$\times$. However, the overall performance is limited by computational steps that cannot be accelerated by the new concurrent propagation strategy. These non-accelerable components include the initial application of multiple Fermi-Dirac operators and frequent matrix-vector inner products involving the current density operator, an operation termed \texttt{xmy} ($\langle \psi_1|\hat{J}_\alpha|\psi_2\rangle$).
As described by Amdahl's law (Eq.~\ref{eq:amdahl}), these non-accelerable components fundamentally limit the achievable overall speedup.
%

%

\begin{figure}[htbp]
    \centering
    \begin{overpic}[width=0.75\columnwidth]{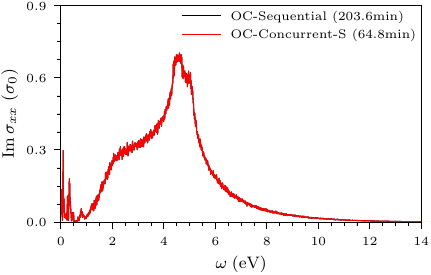}
        \put(-2,57){\bfseries(a)}
    \end{overpic}
    
    \vspace{2.ex} 
    
    \begin{overpic}[width=0.75\columnwidth]{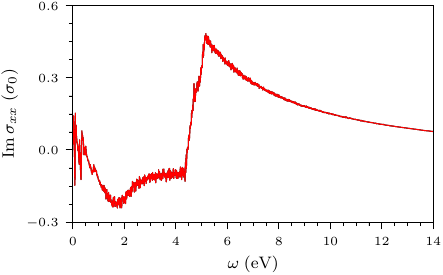}
        \put(-2,57){\bfseries(b)}
    \end{overpic}
    \caption{
    \textbf{Accuracy validation of the new concurrent method for the optical conductivity calculation.}
    The optical conductivity (OC) calculated by the OC-Concurrent-S method (red line) is compared against the baseline OC-Sequential method (black line) for (a) the real part and (b) the imaginary part of the spectrum. The calculation was performed on a magic-angle twisted bilayer graphene supercell containing 4,763,200 atoms ($D_H=120$, $N_t=4096$). The wall time for each method is provided in parentheses within the legend. The optical conductivity is in units of $\sigma_0=e^2/\hbar$.}
    \label{fig:acc_oc}
\end{figure}

The numerical consistency among different implementations of the OC calculation is evaluated in Fig.~\ref{fig:acc_oc}. The infinity norm error analysis reveals $\varepsilon_{\infty}$ values under $2.5 \times 10^{-15}$ for the correlation function $C^{\mathrm{OC}}_{\alpha\beta}(t)$ and below $8.6 \times 10^{-12}$ for the optical conductivity $\sigma_{\alpha\beta}(\hbar\omega)$, attesting to the precision of the methods.

\begin{figure*}[htbp]
    \centering
    \begin{overpic}[width=0.33\textwidth]{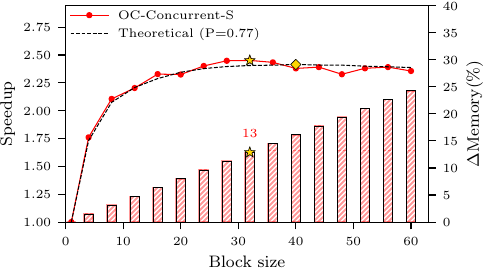}
        \put(-5,50){\bfseries(a)}
    \end{overpic}
    \hfill
    \begin{overpic}[width=0.315\textwidth]{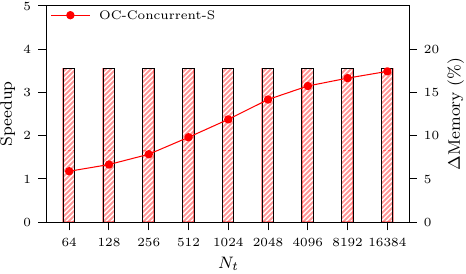}
        \put(-5,52.5){\bfseries(b)}
    \end{overpic}
    \hfill
    \begin{overpic}[width=0.32\textwidth]{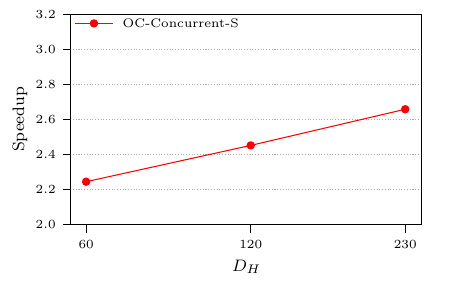}
        \put(-5,51.5){\bfseries(c)}
    \end{overpic}

  \caption{%
  \textbf{Performance comparison of the OC-Concurrent-S method vs. the baseline OC-Sequential method.}
    (a) Speedup (line) and relative memory consumption (bars) as a function of block size $b$. Asterisks and diamonds mark the empirically measured and theoretical optimal values, respectively ($D_H=120$, $N_t=1024$).
    (b) Speedup (line) and relative memory consumption (bars) at the optimal block size $b$, as a function of the total number of time steps $N_t$ ($D_H=120$).
    (c) Speedup at the optimal block size $b$, as a function of the matrix density $D_H$ ($N_t=1024$).
  }
    \label{fig:OC_combined}
\end{figure*}

The computational performance of the new OC-Concurrent-S method is shown in detail in Fig.~\ref{fig:OC_combined}.
First, we examine the performance as a function of the block size $b$, shown in Fig.~\ref{fig:OC_combined}(a). The speedup exhibits a clear peak behavior (first increasing and then decreasing) which allows us to fit the accelerable fraction of the computation, yielding $P \approx 0.77$. The empirically measured optimal block size is found at $b^*=32$, which provides a speedup of approximately 2.5$\times$ with a manageable memory overhead of about +13\%. This is in close proximity to the theoretical optimum of $b_{\mathrm{opt}}=40$, and as the figure indicates, the performance peak is quite broad, with both block sizes yielding very similar speedups.

Using the optimal block size, we then evaluated the performance as a function of the total number of time steps, $N_t$. As shown in Fig.~\ref{fig:OC_combined}(b),  the speedup increases from approximately 1.2$\times$ to 3.5$\times$ as $N_t$ grows. This behavior is consistent with Amdahl's law (Eq.~\ref{eq:amdahl}): as the total propagation duration increases, the computational share of the accelerable part rises (i.e., the parameter $P$ approaches 1), causing the overall speedup $S$ to approach the propagation-part speedup $n$. Throughout this test, the relative memory overhead for OC-Concurrent-S remains nearly constant at approximately +18\%.

Finally, the impact of the Hamiltonian matrix density $D_H$ is shown in Fig.~\ref{fig:OC_combined}(c). The speedup improves for denser matrices, increasing from approximately 2.2$\times$ to 2.7$\times$ as $D_H$ grows. This trend is consistent with our theoretical analysis: a larger number of non-zero elements increases the $\texttt{amxsy}/\texttt{axpy}$ time ratio, thereby improving the overall speedup.

In summary, for the OC calculation, the OC-Concurrent-S method achieves a robust speedup of 2--3$\times$ at the cost of a modest and predictable memory overhead (13--18\%).

%% file: context/DP.tex
Dynamical polarization (DP) is a key physical quantity for understanding the optical and excitation properties of materials, as it is closely related to the dielectric response and plasmons. According to the Kubo formula~\cite{1957-kubo}, the dynamical polarization is defined as~\cite{2012-yuan-screening,2011-yuan-plasmons}
\begin{equation}
\Pi(\mathbf{q}, \hbar \omega) = -\frac{2}{A} \int_0^\infty \mathrm{e}^{\mathrm{i}\omega t} \mathrm{Im}\,C^{\mathrm{DP}}(t,\mathbf{q})\,\mathrm{d}t,
\end{equation}
where the DP correlation function is given by
\begin{equation}
C^{\mathrm{DP}}(t,\mathbf{q}) = \bigl\langle \psi_1^{\mathrm{DP}}(t) \bigm| \hat\rho(\mathbf{q}) \bigm| \psi_2^{\mathrm{DP}}(\mathbf{q},t)\bigr\rangle.
\end{equation}

The two time-dependent bra and ket in this expression are given by
\begin{align}
\ket{\psi_1^{\mathrm{DP}}(t)} &= \mathrm{e}^{-\mathrm{i}\hat H t}\,f(\hat H)\,\ket{\psi},\\
\ket{\psi_2^{\mathrm{DP}}(\mathbf{q},t)} &= \mathrm{e}^{-\mathrm{i}\hat H t}\,\bigl[1 - f(\hat H)\bigr]\,\hat\rho(-\mathbf{q})\,\ket{\psi}.
\end{align}

The density operator is defined as
\begin{equation}
\hat\rho(\mathbf{q}) =  \sum_{i=1}^{N_{\mathrm{basis}}} \mathrm{e}^{\mathrm{i}\mathbf{q}\cdot\hat{\mathbf{r}}_i}.
\end{equation}

For the numerical implementation, the spectral width is set analogously to the OC calculation, resulting in a rescaled time step of $\tilde{\tau}_{\mathrm{DP}}=\pi/2$. The conventional sequential sPM, which we term the DP-Sequential method, computes the correlation function by sequentially propagating both states at each time step.

The DP calculation is accelerated using our general-purpose state-based implementation (DP-Concurrent-S).
Similar to the OC calculation, the performance of the DP-Concurrent-S method is limited by its non-accelerable components: the application of multiple Fermi-Dirac operators and frequent matrix-vector inner products. However, the \texttt{xmy} inner product in the DP calculation, which involves the density operator ($\langle \psi_1|\hat{\rho}(\mathbf{q})|\psi_2\rangle$), is computationally much faster than its counterpart involving the current density operator in the OC case. This crucial difference results in a larger accelerable fraction $P$ for the DP calculation within the Amdahl's law framework, allowing for a higher overall speedup.

\begin{figure}[htbp]
    \centering
    \includegraphics[width=0.8\columnwidth]{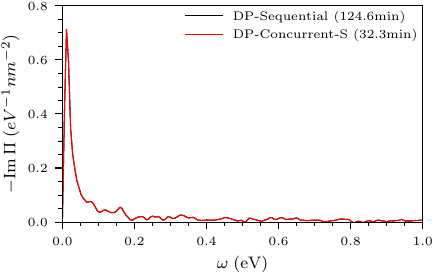}
    \caption{
    \textbf{Accuracy validation of the new concurrent method for the dynamical polarization calculation.} 
    The imaginary parts of the dynamical polarization (DP) calculated by the DP-Concurrent-S method (red line) are compared against the baseline DP-Sequential method (black line). The calculation was performed on a magic-angle twisted bilayer graphene supercell containing 4,763,200 atoms ($D_H=120$, $N_t=4096$). The wall time for each method is provided in parentheses within the legend.}
    \label{fig:acc_dp}
\end{figure}

Validation of the DP-Concurrent-S method, as illustrated in Fig.~\ref{fig:acc_dp}, demonstrates excellent agreement with the baseline DP-Sequential calculation. The corresponding infinity norm errors are $3.5 \times 10^{-16}$ for the correlation function $C^{\mathrm{DP}}(t,\mathbf{q})$ and $2.6 \times 10^{-12}$ for the dynamical polarization $\Pi(\mathbf{q}, \hbar \omega)$, underscoring the reliability of the proposed approach.

\begin{figure*}[htbp]
    \centering
    \begin{overpic}[width=0.33\textwidth]{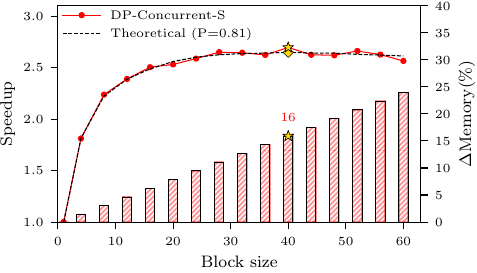}
        \put(-5,50){\bfseries(a)}
    \end{overpic}
    \hfill
    \begin{overpic}[width=0.315\textwidth]{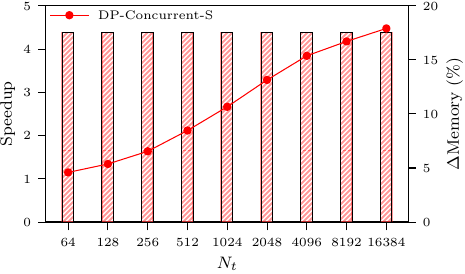}
        \put(-5,52.5){\bfseries(b)}
    \end{overpic}
    \hfill
    \begin{overpic}[width=0.32\textwidth]{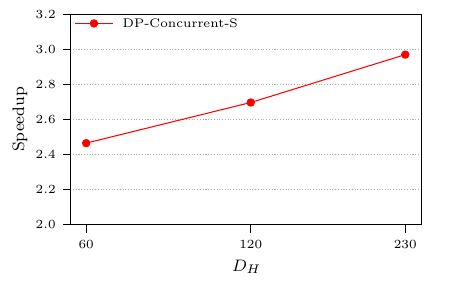}
        \put(-5,51.5){\bfseries(c)}
    \end{overpic}

  \caption{%
  \textbf{Performance comparison of the DP-Concurrent-S method vs. the baseline DP-Sequential method.}
    (a) Speedup (line) and relative memory consumption (bars) as a function of block size $b$. Asterisks and diamonds mark the empirically measured and theoretical optimal values, respectively ($D_H=120$, $N_t=1024$).
    (b) Speedup (line) and relative memory consumption (bars) at the optimal block size $b$, as a function of the total number of time steps $N_t$ ($D_H=120$).
    (c) Speedup at the optimal block size $b$, as a function of the matrix density $D_H$ ($N_t=1024$).
  }
    \label{fig:DP_combined}
\end{figure*}

The computational performance of the DP-Concurrent-S method was benchmarked following the same protocols established for the OC calculation. As shown in Fig.~\ref{fig:DP_combined}, the speedup exhibits a peak at the theoretically predicted and empirically confirmed optimal block size of $b^*=40$, which yields a speedup of approximately 2.7$\times$ with a manageable memory overhead of +16\%. As expected, the speedup increases from approximately 1.1$\times$ to 4.5$\times$ with $N_t$, while the memory cost remains constant at approximately +18\%. Furthermore, the performance improves with increasing matrix density $D_H$, with the speedup growing from approximately 2.5$\times$ to 3.0$\times$. This robust performance, which is slightly better than that for the OC calculation, is a direct result of the faster density operator inner product.

In summary, the DP-Concurrent-S method provides a reliable and efficient approach for calculating dynamical polarization, achieving a speedup of nearly 3$\times$ with a modest and predictable memory increase  (13--18\%).

%% file: context/CD.tex

Charge density (CD) is a fundamental quantity in electronic‐structure calculations, particularly in density functional theory. In the sPM, the charge density is computed as~\cite{2023-dfpm}
\begin{equation}
\rho(\mathbf{r}) = \frac{N_{\mathrm{basis}}}{2\pi} \int_{-\infty}^{\infty} \Bigl|\bra{\mathbf{r}}\mathrm{e}^{-\mathrm{i}\hat H t}f^{1/2}(\hat H)\ket{\psi}\Bigr|^2 \,\mathrm{d}t,
\label{eq:rho}
\end{equation}
where $N_{\mathrm{basis}}$ is the total number of basis functions. The Fermi–Dirac operator $f^{1/2}(\hat{H})$ filters the random state $\ket{\psi}$, such that the resulting state contains information only within the occupied spectral width, $\Delta_{\mathrm{occ}}$. The modulus‐square operation then doubles the spectral span to an effective width of $2\Delta_{\mathrm{occ}}$. 
Consequently, the time step is chosen as $\tau_{\mathrm{CD}} = \pi/\Delta_{\mathrm{occ}}$, which corresponds to a rescaled time step of $\tilde{\tau}_{\mathrm{CD}} = \pi|H|/\Delta_{\mathrm{occ}}$. The conventional sequential sPM, which we term the CD-Sequential method, computes this quantity by sequentially propagating the state.

The CD calculation can be accelerated using our general-purpose state-based implementation (CD-Concurrent-S).
Similar to the OC and DP calculations, its performance is limited by non-accelerable components. However, the non-accelerable workload for CD calculation is significantly smaller for two reasons. First, the CD calculation requires fewer applications of the Fermi-Dirac operator. Second, another non-accelerable component of the CD calculation, the element-wise operation \texttt{abs2py} ($z_i=|x_i|^2+y_i$), is computationally less expensive than the \texttt{xmy} inner product used in the OC and DP cases. This combined reduction in the non-accelerable cost leads to a larger accelerable fraction $P$ for the CD calculation within the Amdahl's law framework, allowing for a higher potential speedup.
However, a more fundamental constraint arises from the rescaled time step, $\tilde{\tau}_{\mathrm{CD}}$. 
This value is inversely proportional to the ratio of the occupied spectral width to the total spectral width, which is equivalent to the ratio of the number of electrons to the basis set size. For systems with large basis sets, this ratio is small, leading to a large $\tilde{\tau}_{\mathrm{CD}}$. A large time step pushes the baseline method into a regime where the expansion load is already low, thus diminishing the potential speedup offered by the new method.

\begin{figure}[htbp]
    \centering
    \includegraphics[width=0.8\columnwidth]{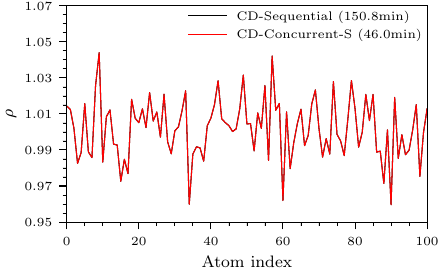}
    \caption{
    \textbf{Accuracy validation of the new concurrent method for the charge density calculation.} 
    The charge density (CD) calculated by the CD-Concurrent-S method (red line) are compared against the baseline CD-Sequential method (black line). The calculation was performed on a magic-angle twisted bilayer graphene supercell containing 4,763,200 atoms ($D_H=120$, $N_t=4096$). The wall time for each method is provided in parentheses within the legend.}
    \label{fig:acc_cd}
\end{figure}

We next validate the accuracy of the CD-Concurrent-S method. As shown in Fig.~\ref{fig:acc_cd}, the result is in excellent agreement with the baseline result. The quantitative infinity norm error ($\varepsilon_{\infty}$) for the charge density $\rho(\mathbf{r})$ is less than $1.1 \times 10^{-12}$, confirming the high numerical accuracy of the new method.

\begin{figure*}[htbp]
    \centering
    \begin{overpic}[width=0.33\textwidth]{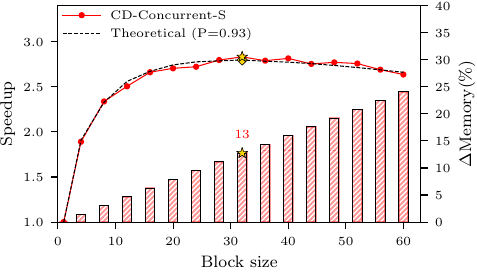}
        \put(-5,50){\bfseries(a)}
    \end{overpic}
    \hfill
    \begin{overpic}[width=0.315\textwidth]{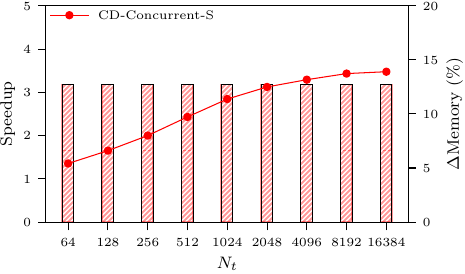}
        \put(-5,52.5){\bfseries(b)}
    \end{overpic}
    \hfill
    \begin{overpic}[width=0.32\textwidth]{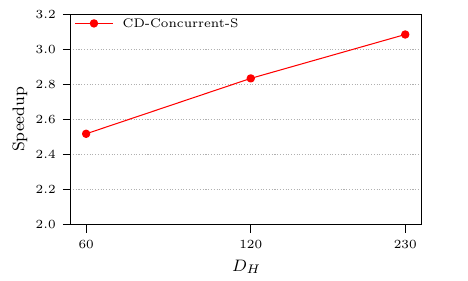}
        \put(-5,51.5){\bfseries(c)}
    \end{overpic}

  \caption{%
  \textbf{Performance comparison of the CD-Concurrent-S method vs. the baseline CD-Sequential method.}
    (a) Speedup (line) and relative memory consumption (bars) as a function of block size $b$. Asterisks and diamonds mark the empirically measured and theoretical optimal values, respectively ($D_H=120$, $N_t=1024$).
    (b) Speedup (line) and relative memory consumption (bars) at the optimal block size $b$, as a function of the total number of time steps $N_t$ ($D_H=120$).
    (c) Speedup at the optimal block size $b$, as a function of the matrix density $D_H$ ($N_t=1024$).
  }
    \label{fig:CD_combined}
\end{figure*}

The computational performance of the CD-Concurrent-S method was evaluated using the established protocols for the OC and DP calculations, with the results shown in Fig.~\ref{fig:CD_combined}. The speedup as a function of block size $b$ exhibits the familiar peak behavior, yielding a fitted accelerable fraction of $P \approx 0.93$. The optimal performance is achieved at a block size of $b^*=32$ (which matches theory), providing a speedup of approximately 2.8$\times$ with a memory overhead of +13\%. And the speedup increases from approximately 1.4$\times$ to 3.5$\times$ with $N_t$, while the memory cost remains constant at approximately +13\%. This rapid increase, the fastest among the three state-based implementations, is a direct result of the smaller non-accelerable workload in the CD calculation. Similarly, the performance improves for denser matrices, with the speedup increasing from approximately 2.5$\times$ to 3.1$\times$ as $D_H$ grows.

While the performance is robust for the system tested, it is worth noting that in the CD calculation, the rescaled time step depends on the ratio of the occupied spectral width to the total spectral width. In the present test, which considers only $p_z$ orbitals, the occupied spectrum is exactly half of the total spectral width, resulting in $\tilde \tau_{CD} = \pi$. In calculations with large basis sets, this time step would increase as the fraction of occupied states decreases, which would in turn reduce the achievable speedup.

In summary, for systems with a high ratio of occupied states to basis sets, the CD-Concurrent-S method is highly efficient, offering a speedup of nearly 3$\times$ with a modest memory cost(approximately +13\%).
Generally speaking, for the three physical quantities that rely exclusively on the state-based implementation (OC, DP, and CD), the new method can achieve a speedup of around 3$\times$ at the cost of a small memory overhead ($<20\%$). Furthermore, the speedup automatically improves with increasing $D_H$, making the method well-suited for treating complex systems with strong interactions.

%% file: context/conclusion.tex
%
This work introduces an efficient propagation strategy that bypasses conventional sequential computation while respecting the Nyquist-Shannon sampling theorem. By leveraging a single long-time propagation, our approach reconstructs all intermediate states via a linear combination of shared Chebyshev states, eliminating step-by-step evolution. We developed three tailored implementations—state-, moment-, and energy-based—for different computational targets, alongside a time-blocking strategy to balance efficiency and memory. The method demonstrates broad universality in linear-scaling sPM, enabling over an order-of-magnitude acceleration in property calculations.

We implemented these techniques for the calculation of various properties, including but not limited to the density of states, local density of states, quasi-eigenstates, electronic conductivity, optical conductivity, dynamical polarization, and charge density, and performed systematic numerical benchmarks on a magic-angle twisted bilayer graphene with billions of atoms. While the results presented here are for tight-binding models, the concurrent methodology is directly applicable to other frameworks, including density functional theory implemented with orthogonal basis sets.
Furthermore, all our benchmarks show that the speedup is more significant for denser matrices (i.e., larger $D_H$), indicating that the method is particularly well-suited for treating complex systems with strong interactions.


The concurrent stochastic propagation method presented here offers a transformative increase in computational efficiency for large-scale systems. It achieves order-of-magnitude speedups—reducing simulation times from days to hours for billion-atom models—without incurring memory overhead or sacrificing precision. Furthermore, its fundamental departure from sequential time-evolution provides a general framework for optimizing a broad class of computational algorithms.

%% file: context/cost_mem_table.tex
Tab.~\ref{tab:cost} provides a quantitative comparison of the primary mathematical operations and memory requirements for the various algorithms discussed in the main text. For clarity, the table focuses on operations whose costs depend on the tunable block size $b$ and omits several whose costs are independent of it. Specifically, the omitted operations include the matrix-vector inner product \texttt{xmy} = $\langle x|M|y\rangle$ for the optical conductivity and dynamical polarization calculations, and the element-wise operation \texttt{abs2py} (where $z_i=|x_i|^2 + y_i$) for the charge density calculation.

\begin{table*}[htbp]
\begin{threeparttable}

\caption{Comparison of the number of primary mathematical operations and memory usage for algorithms propagating over a fixed total duration $t_{\mathrm{tot}} = N_t \cdot \tau$. Time points are denoted by $t_j = j \cdot \tau$, where $j$ may be a global index or a local index within a time block; $d$ is the system dimensionality. 
In the new method, each block comprises $b$ time steps (duration $t_b = b \cdot \tau$) and there are $N_b = N_t / b$ blocks. $N_q$ indicates the number of $\mathbf{q}$ points. 
The \texttt{states} column reports the maximum number of states that must be stored concurrently during propagation.} 
\label{tab:cost}
\begin{ruledtabular}
    \begin{tabular}{lcccccc}
    Algorithm 
    & \# \texttt{amxsy}
    & \# \texttt{axpy}
    & \# \texttt{dot}
    & \# \texttt{states} 
    & Tunable\\
    \midrule
    DOS-Sequential &
    $\tilde{t}_{\mathrm{tot}} \cdot R(\tilde{\tau})$ &
    $N_{t}\cdot N(\tilde{\tau})$ &
    $N_{t}$& $4$ & ---  \\
    DOS-Concurrent-M & 
    $\tilde{t}_{\mathrm{tot}} \cdot R(\tilde{t}_{\mathrm{tot}})/2$ &
    --- &
    $N(\tilde{t}_{\mathrm{tot}})$& $3$ & ---\\
    \midrule
    QE-Sequential &
    $2\tilde{t}_{\mathrm{tot}} \cdot R(\tilde{\tau})$ &
    $2N_{t}\cdot N(\tilde{\tau})+2N_{t}\cdot N_E$  &
    --- & $N_E+6$ & $N_E$ \\
    QE-Concurrent-S &
    $2\tilde{t}_{\mathrm{tot}} \cdot R(\tilde{t}_b)$ &
    $2N_{b}\cdot \sum_{j=1}^{b} N(\tilde{t}_j)+2N_{t}\cdot N_E$&
    --- & $N_E+2b+4$  & $N_E$; $b$ \\
    QE-Concurrent-E &
    $\tilde{t}_{\mathrm{tot}} \cdot R(\tilde{t}_{\mathrm{tot}})$ &
    $N(\tilde{t}_{\mathrm{tot}})\cdot N_E$ &
    --- & $N_E+3$ & $N_E$  \\
    \midrule
    EC-Sequential &
    $d\cdot\tilde{t}_{\mathrm{tot}} \cdot R(\tilde{\tau})$ & 
    $d\cdot N_{t}\cdot N(\tilde{\tau})$ &
    $N_E(1+d\cdot N_{t})$ & 
    $d\cdot N_{E}+d+4$ & $N_E$\\
    
    EC-Concurrent-S &
    $d\cdot\tilde{t}_{\mathrm{tot}} \cdot R(\tilde{t}_b)$ & 
    $d\cdot N_{b}\cdot \sum_{j=1}^{b} N(\tilde{t}_j)$ &
    $N_E(1+d\cdot N_{t})$ & 
    $d\cdot N_{E}+d+b+3$ & $N_E$; $b$\\
    
    EC-Concurrent-M &
    $d\cdot\tilde{t}_{\mathrm{tot}} \cdot R(\tilde{t}_{\mathrm{tot}})$ & 
    --- &
    $d\cdot N_E \cdot N(\tilde{t}_{\mathrm{tot}})$ & 
    $d\cdot N_{E}+d+3$ & $N_E$\\
    \midrule
    OC-Sequential &
    $(d+1)\cdot\tilde{t}_{\mathrm{tot}} \cdot R(\tilde{\tau})$ & 
    $(d+1)\cdot N_{t}\cdot N(\tilde{\tau})$  &
    --- &
     $d+6$ & ---\\
    
    OC-Concurrent-S &
    $(d+1)\cdot\tilde{t}_{\mathrm{tot}} \cdot R(\tilde{t}_b)$ & 
    $(d+1)\cdot N_{b}\cdot \sum_{j=1}^{b} N(\tilde{t}_j)$ &
    --- &
     $d+2b+4$ & $b$\\
    \midrule
    DP-Sequential &
    $2N_q\cdot\tilde{t}_{\mathrm{tot}} \cdot R(\tilde{\tau})$ & 
    $2N_q\cdot N_{t} \cdot N(\tilde{\tau})$ &
    ---&  $8$ & $N_q$\\
    
    DP-Concurrent-S &
    $2N_q\cdot\tilde{t}_{\mathrm{tot}} \cdot R(\tilde{t}_b)$ & 
    $2N_q\cdot N_{b}\cdot \sum_{j=1}^{b} N(\tilde{t}_j)$ &
    --- & $2b+6$ & $b$; $N_q$\\
    \midrule
    CD-Sequential &
    $2\tilde{t}_{\mathrm{tot}} \cdot R(\tilde{\tau})$ & 
    $2N_{t}\cdot N(\tilde{\tau})$ &
    ---& $7$ & ---\\
    
    CD-Concurrent-S &
    $2\tilde{t}_{\mathrm{tot}} \cdot R(\tilde{t}_b)$ & 
    $2N_{b}\cdot \sum_{j=1}^{b} N(\tilde{t}_j)$ &
    ---& $2b+5$ & $b$\\
    \end{tabular}
\end{ruledtabular}

\end{threeparttable}
\end{table*}

%% file: context/poly_expansions.tex
The expansion in Chebyshev polynomials of the first kind, $T_k(\tilde{H})$, is a powerful numerical method, renowned for its efficiency and numerical stability. These polynomials are generated via the well-known three-term recurrence relation, with the initial cases being $T_0(\tilde{H})=1$ and $T_1(\tilde{H})=\tilde{H}$:
\begin{equation}
    T_{k+1}(\tilde{H})=2\tilde{H} T_{k}(\tilde{H})-T_{k-1}(\tilde{H}).
    \label{eq:cheb_relation}
\end{equation}
These polynomials are particularly effective for expanding functions. When expanding the time-evolution operator $\mathrm{e}^{-\mathrm{i}\tilde{H}\tilde{\tau}}$ in a series of Chebyshev polynomials, the expansion coefficients, $c_n(\tilde{\tau})$, are given by:
\begin{equation}
c_n(\tilde{\tau}) = \begin{cases}
    J_0(\tilde{\tau}) & n = 0 \\
    2(-\mathrm{i})^n J_n(\tilde{\tau}) & n \ge 1
\end{cases}
\end{equation}
where $J_n(\tilde{\tau})$ is the Bessel function of the first kind of integer order $n$.

%
The Jacobi polynomial $P_n^{(\alpha,\beta)}(x)$~\cite{2011-qin-poly,2025-poly} expansion method imposes no special requirements on the form of the Hamiltonian and is applicable to arbitrarily complex spectral structures. The parameters $\alpha$ and $\beta$ can also be tuned according to the form of the Hamiltonian to achieve optimal performance. However, this method requires that the spectrum of the Hamiltonian matrix be rescaled to the interval $[-1,1]$. The weight function for the Jacobi polynomials $P_n^{(\alpha,\beta)}(x)$ is
\begin{equation}
   w(x) = (1-x)^{\alpha}(1+x)^{\beta},
\end{equation}
and the recurrence relation is given by~\cite{2025-poly}
\begin{equation}
\begin{aligned}
P_{0}^{(\alpha,\beta)}(x) &= 1,\\
P_{1}^{(\alpha,\beta)}(x) &= a_0x+b_0,\\
    P_{n+1}^{(\alpha,\beta)}(x) &= \bigl(a_n x + b_n\bigr)\,P_{n}^{(\alpha,\beta)}(x)
- c_n\,P_{n-1}^{(\alpha,\beta)}(x),
\end{aligned}
\label{eq:jacobi_rec}
\end{equation}
where the coefficients are:
\begin{equation}
\begin{aligned}
a_n &= \frac{(2n+\alpha+\beta+1)(2n+\alpha+\beta+2)}
             {2\,(n+1)(n+\alpha+\beta+1)},\\
b_n &= \frac{(2n+\alpha+\beta+1)(\alpha^2-\beta^2)}
             {2\,(n+1)(n+\alpha+\beta+1)(2n+\alpha+\beta)},\\
c_n &= \frac{(n+\alpha)(n+\beta)(2n+\alpha+\beta+2)}
             {(n+1)(n+\alpha+\beta+1)(2n+\alpha+\beta)}.
\end{aligned}
\end{equation}
By tuning the parameters $\alpha$ and $\beta$ in the Jacobi polynomials $P_n^{(\alpha,\beta)}(x)$, various other orthogonal polynomials can be obtained. In particular, for $\alpha=\beta=0$, they are equivalent to the Legendre polynomials $P_n(x)$. For $\alpha=\pm 1/2$ and $\beta=\pm 1/2$, they reduce to the Chebyshev polynomials of the first to the fourth kind, and the recurrence relation still satisfies Eq.~\ref{eq:jacobi_rec}.

The generalized Laguerre polynomials $L_n^{\alpha}(x)$~\cite{2011-spectral}, with a weight function of the form $\mathrm{e}^{-x}x^{\alpha}$, are more suitable for Hamiltonian spectra that are distributed over the interval $[0,\infty)$ and exhibit an exponential decay profile. Examples include descriptions of hydrogen atom wave functions and quantum scattering problems. The commonly used recurrence relation for $L_n^{\alpha}(x)$ is:
\begin{equation}
\begin{aligned}
L_{0}^{(\alpha)}(x) &= 1,\\
L_{1}^{(\alpha)}(x) &= 1+ \alpha -x,\\
L_{n+1}^{(\alpha)}(x) &=\frac{2n+\alpha+1 - x}{n+1}\,L_{n}^{(\alpha)}(x)
-\frac{n+\alpha}{n+1}\,L_{n-1}^{(\alpha)}(x).
\end{aligned}
\end{equation}

The Hermite polynomials $H_n(x)$~\cite{2010-qin-poly,2011-qin-poly}, with a weight function of the form $\mathrm{e}^{-x^2}$, are more suitable for Hamiltonian spectra that have a Gaussian profile over the entire real axis. They are often used in calculations for the quantum harmonic oscillator. The recurrence relation for $H_n(x)$ is:
\begin{equation}
\begin{aligned}
H_{0}(x) &= 1,\\
H_{1}(x) &= 2x,\\
H_{n+1}(x) &= 2x\,H_n(x)-2n\,H_{n-1}(x).
\end{aligned}
\end{equation}

When expanding an arbitrary function using the aforementioned orthogonal polynomials, the coefficient for each polynomial term can be obtained by leveraging their orthogonality after the Hamiltonian spectrum has been rescaled to the appropriate interval. However, the structure of the time-evolution operator $\mathrm{e}^{\mathrm{i}\Hamil \tau}$ does not align well with the properties of the generalized Laguerre and Hermite polynomials; therefore, we do not benchmark these two expansion methods. As noted in the main text, the coefficients for the expansion of the time-evolution operator $\mathrm{e}^{-\mathrm{i}\tilde{H} \tilde{\tau}}$ in Chebyshev polynomials are related to the Bessel functions $J_n(\tilde{\tau})$. Similarly, for an expansion in Legendre polynomials, the coefficients are related to the spherical Bessel functions $j_n(\tilde{\tau})$, and the expansion is given by:
\begin{equation}
    \mathrm{e}^{-\mathrm{i}\tilde{H} \tilde{\tau}}
=\sum_{n=0}^{N(\tilde{\tau})} (2n+1)\,\mathrm{i}^n\,j_n(\tilde{\tau})\;P_n\!\bigl(\widetilde H\bigr).
\end{equation}

In addition to orthogonal polynomials, we also tested the classic Taylor series expansion for the time-evolution operator $\mathrm{e}^{-\mathrm{i}\tilde{H} \tilde{\tau}}$:
\begin{equation}
\mathrm{e}^{-\mathrm{i}\tilde{H} \tilde{\tau}} = \sum_{n=0}^{N(\tilde{\tau})} \frac{(\mathrm{i}\tilde{\tau})^n}{n!}\tilde{H} ^n.
 \label{eq:taylor}
\end{equation}
The Taylor expansion is highly effective for small time steps, but as the time step increases, it becomes slow and numerically unstable due to the factorial term in the denominator. The expansion coefficients $(\mathrm{i}\tilde{\tau})^n/n!$ first increase with $n$ to a very large value before rapidly decreasing. In our tests, using a precision threshold of $\eta = 1 \times 10^{-14}$, the coefficients suffered from numerical overflow for expansion orders $n > 200$, making the calculation infeasible.